\documentclass[aps,prl,twocolumn,showpacs,groupaddress]{revtex4-1}
\usepackage{amsmath}
\usepackage{amsfonts}
\usepackage{multirow}
\usepackage{float}
\usepackage{graphicx}
\usepackage{subfig}
\usepackage{nicefrac}
\usepackage{xcolor}
\usepackage{bm}

\usepackage{tikz-cd}
\usepackage[breaklinks=true,pdfborder={0 0 0}]{hyperref}
\hypersetup{colorlinks=true,citecolor=red}

\newcommand{\bra}[1]{\langle #1 \vert}
\newcommand{\ket}[1]{\vert #1 \rangle}

\begin{document}

\title{Phenomenological Ehrenfest Dynamics with Topological and Geometric Phase Effects and the curious case of Elliptical intersection}

\author{Dhruv Sharma}
\affiliation{Department of Physics and Materials Science, University of Luxembourg, L-1511 Luxembourg City, Luxembourg}
\date{\today}

\begin{abstract}
We present a comprehensive computational framework for simulating nonadiabatic molecular dynamics with explicit inclusion of geometric phase (GP) effects. Our approach is based on a generalized two-level Hamiltonian model that can represent various electronic state crossings—conical intersections, avoided crossings, and elliptic intersections—through appropriate parameterization. We introduce a novel prelooping trajectory initialization scheme, allowing us to encode the memory as an initial phase accumulated due to the adiabatic evolution over the potential energy surface. This is a unified framework to handle different types of level crossings by incorporating Berry curvature-based force corrections to Ehrenfest dynamics, ensuring accurate representation of topological effects. For conical intersections, our method incorporates the theoretically expected phase $\pi$, while for elliptic intersections, it yields a parametrically tunable but loop-radius (energy) independent phase different from $\pi$. We also include an eccentricity parameter ($e$) in the diabatic coupling to model more realistic molecular systems. Numerical simulations demonstrate the consistency of our approach with theoretical predictions for mixing of states and inhibition from mixing due to geometric phase effects. This framework provides a valuable tool for studying quantum-classical interactions in molecular systems where geometric phase effects play a significant role. The elliptical intersection and geometric phase effect opens avenue for the design and discovery of degenerate materials. It produces a fresh look to help develop a new kind of spectroscopy and potential qubit applications. This simple Hamiltonian reveals a pathological phase protection effect $(E=\kappa*r, \kappa \in \mathbb{R})$ that has great utility in a new spectroscopy design.

\end{abstract}

\maketitle

\section{Introduction}

Nonadiabatic molecular dynamics is essential for understanding processes where electronic and nuclear motions are strongly coupled, such as photochemical reactions, energy transfer in biological systems, and charge transport in materials. Nonadiabatic molecular dynamics is crucial for understanding processes where the Born-Oppenheimer approximation fails, particularly near conical intersections (CIs) where electronic potential energy surfaces are degenerate.\cite{Liu2023Nonadiabatic,Nelson2014Nonadiabatic, Nelson2020Non-adiabatic,intersections2011theory}. CIs are points where two or more electronic potential energy surfaces become degenerate. They play a central role in photochemistry, facilitating processes like isomerization, photodissociation, and internal conversion, and are now recognized as a dominant mechanism for coupled charge and vibrational energy flow in excited states \cite{Yarkony1996Diabolical,Schuurman2018Dynamics,Domcke2012Role,Worth2004Beyond}. CIs also influence energy transfer and charge transport by providing pathways for rapid electronic relaxation \cite{Matsika2021Electronic,An2020Nonadiabatic}. Passage through a CI converts electronic energy into vibrational energy, driving photochemical reactions. The direction and velocity of approach to a CI, as well as the topography of the intersection, significantly affect reaction rates, yields, and product distributions \cite{Schuurman2018Dynamics,Farfan2020A,You2015Structure}. Modern electronic structure methods and nonadiabatic molecular dynamics simulations have advanced the ability to characterize and predict CI-mediated processes. New algorithms and computational models allow for efficient exploration of CI topographies and their impact on photochemical observables, even in large or complex systems \cite{Xu2025Conical,Pieri2021The,Ibele2022A,Matsika2021Electronic,Angelico2024Determining}.

At conical intersections, the geometric phase (GP) effect, first described by Berry \cite{Berry1984Quantal}, manifests itself as a topological sign change in the electronic wave function when the nuclear motion encircles the point of degeneracy. This phase change has profound consequences for quantum dynamics, affecting interference patterns, branching ratios, and reaction outcomes \cite{Mead1992, Ryabinkin2017Geometric,Yuan2018Observation, Joubert-Doriol2013Geometric, Xie2017Constructive}. However, standard computational approaches such as Ehrenfest dynamics and surface hopping often neglect GP effects, resulting in potentially inaccurate predictions for systems with strong nonadiabatic coupling \cite{Ryabinkin2017Geometric,Gherib2015Why,Curchod2018Ab,Crespo‐Otero2018Recent,Agostini2019Different}.
\begin{figure}[ht]
    \centering
    \includegraphics[width=0.5\textwidth]{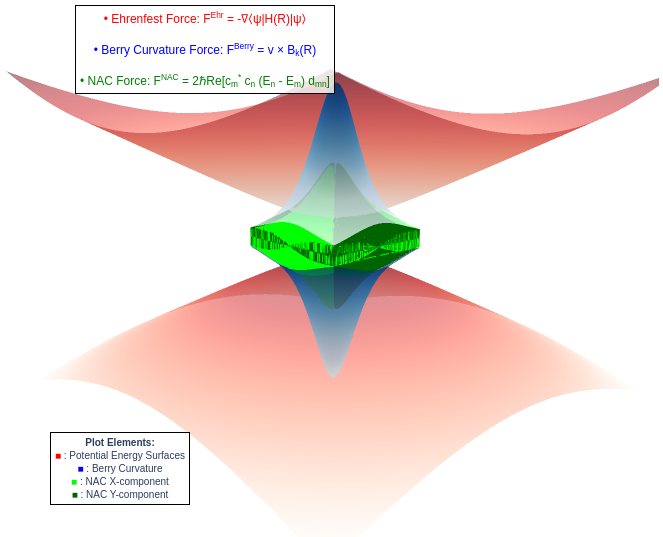}
    \caption{
\textbf{Visualization of quantum-classical force components in a two-level system with tunable intersection.}
Red transparent surfaces show the lower and upper adiabatic potential energy surfaces (PES) as functions of nuclear coordinates $(x, y)$, computed from the model Hamiltonian. Blue surfaces represent the Berry curvature (clipped for visibility) for the lower and upper adiabatic states, highlighting regions of strong geometric phase effects. Light green and dark green lines indicate the real parts of the $x$- and $y$-components of the nonadiabatic coupling (NAC) vector, respectively, computed numerically on a fine grid near the intersection.  This comprehensive visualization illustrates the interplay between mean-field, geometric, and nonadiabatic forces in quantum molecular dynamics near electronic state crossings.
}
    \label{fig:force}
\end{figure}

\begin{figure*}[ht]
\centering
\includegraphics[width=\textwidth]{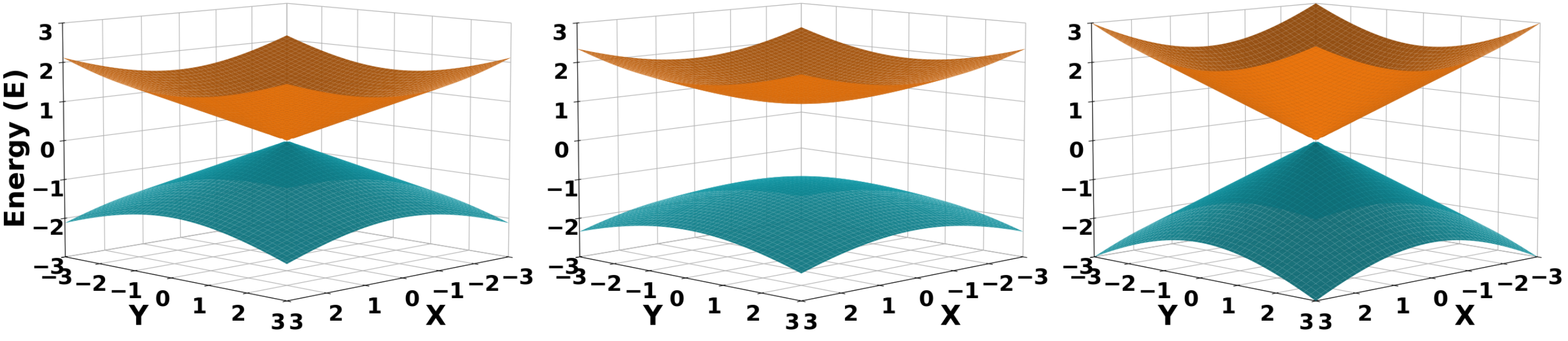}
\caption{Visualization of the different types of electronic state crossings for $a=s=1$ and $e=0$. The adiabatic surfaces of our model Hamiltonian showing how different choices of $Z$ lead to: (a) conical intersection, (b) avoided crossing, and (c) elliptic intersection. The red and blue surfaces represent the upper and lower adiabatic energy surfaces, respectively. Note that while the elliptic intersection appears visually similar to the conical intersection near the crossing point, it exhibits stronger curvature with steeper gradients, spanning a wider energy range (-3 to 3) compared to the conical intersection's narrower range (-2 to 2).}
\label{fig:crossings}
\end{figure*}

 Recent research has focused on understanding when GP effects are significant, how they manifest in molecular observables, and developing new computational strategies to incorporate them into simulations \cite{Althorpe2008Effect,Krotz2022Treating,Li2017Geometric}. Approaches include gauge-invariant derivative couplings, quantum-classical Liouville equation (QCLE) methods, and the use of surface hopping Gaussian phase-space packets, which improve the treatment of quantum coherence and phase effects \cite{Horenko2002Quantum-classical,Krotz2022Treating}. Experimental studies have also validated the importance of including GP in theoretical models, as seen in the $H + HD \rightarrow H_2 + D$ reaction \cite{Yuan2018Observation, Aoiz2020How}. Model system studies have further clarified the impact of GP on observable quantities like scattering cross sections, tunneling lifetimes, and photodissociation dynamics \cite{Wang2025Cold, Lin2018Nonadiabatic, Xie2017Constructive,Mi2023Geometric}.

Ehrenfest dynamics and surface hopping are two widely used mixed quantum-classical methods for simulating nonadiabatic molecular dynamics. Ehrenfest dynamics is computationally efficient but often fails to capture quantum coherence and correct population transfer, while surface hopping improves on these aspects but introduces its own challenges, especially regarding decoherence and detailed balance. Recent advances aim to combine the strengths of both methods and address their limitations. New methods like multiconfigurational Ehrenfest and multiconfigurational surface hopping (MCSH) combine trajectory branching and mean-field effects, improving accuracy in nonadiabatic dynamics and better capturing quantum coherence and population transfer \cite{Li2024Multiconfigurational,Li2022A,Brink2022Real-time, Freixas2021Nonadiabatic}. Augmented Ehrenfest and parameter-free decoherence schemes have been developed to bridge the gap between mean-field and surface hopping, providing more accurate population dynamics without empirical parameters \cite{Subotnik2010Augmented, Esch2021An, Han2024Nonadiabatic, Nijjar2019Ehrenfest}. Surface hopping generally outperforms Ehrenfest in long-time population dynamics, while Ehrenfest is more accurate for ultrashort time coherence. Multiconfigurational methods can approach exact results in small systems but are computationally demanding for larger systems \cite{Li2024Multiconfigurational, Freixas2021Nonadiabatic, Brink2022Real-time}. Machine learning (ML) is increasingly used to accelerate and enhance the simulation of excited state dynamics in molecules \cite{Westermayr2020Machine, mausenberger2024s, barrett2025transferable, dral2021molecular}. ML methods can significantly reduce computational costs and enable simulations of larger systems or longer timescales, while maintaining reasonable accuracy. To address the problem of geometric phase effect in the nuclear part of the wave-function, fully quantum approaches like time-dependent variational principle (TDVP), Exact-factorization (EF) and Quantum hydrodynamics (QHD) has been developed. TDVP approximates quantum dynamics by projecting evolution onto a variational manifold, linking quantum and classical action principles\cite{kramer2008review, joubert2018nonadiabatic, izmaylov2017quantum}. The geometric phase in the EF framework is directly related to observable quantities, such as the circulation of the nuclear momentum field, and can influence measurable properties such as nuclear current and reaction results \cite{Ha2022Independent, Ibele2023On}. The geometric phase in quantum hydrodynamics framework, as studied by Martinazzo and Burghardt, is a dynamic gauge-invariant phase that evolves in real time and can change due to non-conservative electron-nuclear forces, extending beyond the traditional topological view \cite{Martinazzo2023Dynamics}. 

In this work, we present a phenomenological Ehrenfest dynamics approach that explicitly incorporates GP effects through a pre-looping trajectory initialization scheme that ensures efficient sampling of the geometric phase by generating trajectories that have been assumed to have orbited around the crossing point. The force corrections based on the Berry curvature to account for the topological nature of the crossing region further extend the effect during trajectory evolution. Our approach is based on a two-dimensional (2D) Hamiltonian model that can represent various types of electronic state crossing through appropriate parameterization.

\section{Mixed Quantum Classical Dynamics}

We consider a mixed quantum-classical molecular system within the framework of molecular dynamics where electronic degrees of freedom are treated quantum mechanically while nuclear coordinates evolve classically. The system is characterized by a composite phase space $\Gamma = \mathcal{M} \times \mathcal{H}_{\text{el}}$, where $\mathcal{M} \cong \mathbb{R}^4$ denotes the classical nuclear phase space with coordinates $\mathbf{R} = (x, y) \in \mathbb{R}^2$ and conjugate momenta $\mathbf{P} = (p_x, p_y) \in \mathbb{R}^2$, and $\mathcal{H}_{\text{el}} = \mathbb{C}^2$ represents the two-dimensional complex Hilbert space of electronic states.

The total energy functional governing the system dynamics takes the form:
\begin{equation}
E_{\text{total}}[\mathbf{R}, \mathbf{P}, |\psi\rangle] = T_N(\mathbf{P}) + \langle\psi|\hat{H}_{\text{el}}(\mathbf{R})|\psi\rangle
\end{equation}
where $T_N(\mathbf{P}) = \frac{|\mathbf{P}|^2}{2M}$ is the classical nuclear kinetic energy with nuclear mass $M > 0$, and $\hat{H}_{\text{el}}(\mathbf{R}): \mathcal{H}_{\text{el}} \rightarrow \mathcal{H}_{\text{el}}$ is the parametrically $\mathbf{R}$-dependent electronic Hamiltonian operator acting on the normalized electronic state $|\psi\rangle \in \mathcal{H}_{\text{el}}$ with $\langle\psi|\psi\rangle = 1$.

The electronic Hamiltonian $\hat{H}_{\text{el}}(\mathbf{R})$ is a Hermitian operator for each fixed $\mathbf{R} \in \mathbb{R}^2$, ensuring real eigenvalues corresponding to physical energy levels. The parametric dependence on $\mathbf{R}$ encodes the potential energy surfaces and nonadiabatic couplings characteristic of the linear vibronic coupling model, which will be specified in the diabatic representation in the following subsection.

This formulation explicitly separates the classical nuclear phase space evolution from the quantum electronic dynamics, with the coupling between these subsystems mediated through the parametric dependence of $\hat{H}_{\text{el}}$ on the instantaneous nuclear configuration.

\subsection{Electronic Hamiltonian and Linear Vibronic Coupling}

In the diabatic representation, the electronic Hamiltonian takes the matrix form:
\begin{equation}
\hat{H}_{\text{el}}(\mathbf{R}) = \frac{s}{2} \begin{pmatrix} 
aE(x,y) & x - i\sqrt{1-e^2}y \\ 
x + i\sqrt{1-e^2}y & -E(x,y) 
\end{pmatrix}
\end{equation}
where $s > 0$ is an overall energy scaling factor, $a \in \mathbb{R}$ controls the energetic asymmetry between diabatic states, $0 \leq e < 1$ is an eccentricity parameter modulating the coupling anisotropy, and $E(x,y)$ determines the crossing topology.

The off-diagonal coupling term represents a generalized linear vibronic coupling:
\begin{equation}
V_{12} = \frac{s}{2}[x - i\sqrt{1-e^2}y] = \frac{s}{2}[\lambda_x Q_x + i\lambda_y Q_y]
\end{equation}
with coupling constants $\lambda_x = 1$ and $\lambda_y = \sqrt{1-e^2}$, and nuclear coordinates $Q_x = x$, $Q_y = y$. This generalizes the standard $E \otimes e$ Jahn-Teller model $V_{12} = \lambda(Q_x + iQ_y)$ by introducing elliptical anisotropy through the eccentricity parameter $e$.

In the adiabatic representation, the Hamiltonian is diagonalized in each nuclear configuration, resulting in eigenvalues $E_\pm(\mathbf{R})$ and eigenstates $|\psi_\pm(\mathbf{R})\rangle$. For our model, the adiabatic energies are:
\begin{equation}
E_\pm(\mathbf{R}) = \frac{s}{4}(a-1)E \pm \frac{s}{2}\sqrt{\frac{(a+1)^2}{4}E^2 + x^2 + (1-e^2)y^2}.
\end{equation}

The corresponding adiabatic eigenstates can be expressed as:
\begin{subequations}
\label{eq:eigenstates_full}
\begin{align}
    |\psi_-(\mathbf{R})\rangle &= N_- 
        \begin{pmatrix} 
            -(E\alpha - \sqrt{\Delta}) \\
            r\sqrt{\beta}
        \end{pmatrix}, \\
    |\psi_+(\mathbf{R})\rangle &= N_+
        \begin{pmatrix} 
            E\alpha + \sqrt{\Delta} \\
            r\sqrt{\beta}
        \end{pmatrix},
\end{align}
where $\Delta = E^2\alpha^2 + r^2\beta$, and the normalization
factors $N_\pm$ are given by:
\begin{align}
    N_\pm = \frac{1}{\sqrt{r^2\beta + (E\alpha \pm \sqrt{\Delta})^2}}.
\end{align}
\end{subequations}

\subsection{Crossing Topologies}

The function $E(x,y)$ determines the electronic state crossing type:

\textbf{Conical Intersection (CI):} $E(x,y) = 0$ creates a point degeneracy at the origin with Berry phase $\gamma = \pi$.

\textbf{Avoided Crossing (AC):} $E(x,y) = c$ (constant $c > 0$) introduces a finite energy gap, leading to marginal Berry phases.

\textbf{Elliptic Intersection (EI):} $E(x,y) = \sqrt{x^2 + (1-e^2)y^2}$ creates degeneracy along an elliptical seam with tunable geometric phase.

\section{Theoretical Framework}
\subsection{Ehrenfest Dynamics}
Ehrenfest dynamics is a widely used framework in quantum-classical molecular dynamics, where nuclei are treated as classical particles and electrons are treated quantum mechanically. This approach enables the simulation of nonadiabatic processes in complex molecular systems by coupling classical and quantum subsystems \cite{Crespo‐Otero2018Recent, Tully1998Mixed, Parandekar2005Mixed, Castro2013Optimal}. In this approach, the nuclear coordinates $\mathbf{R}(t) = (x(t), y(t))$ evolve according to Newton's equations of motion:

\begin{equation}
M \ddot{\mathbf{R}} = -\nabla_{\mathbf{R}} \langle \psi(t) | \hat{H}(\mathbf{R}) | \psi(t) \rangle,
\end{equation}

where $M$ is the nuclear mass, $\hat{H}(\mathbf{R})$ is the electronic Hamiltonian parameterized by nuclear coordinates, and $|\psi(t)\rangle$ is the time-dependent electronic wavefunction. Concurrently, the electronic state evolves according to the time-dependent Schrödinger equation:

\begin{equation}
i\hbar \frac{d}{dt} |\psi(t)\rangle = \hat{H}(\mathbf{R}(t)) |\psi(t)\rangle.
\end{equation}

For our two-level Hamiltonian model, we can express the electronic wavefunction as a superposition of two states:
\begin{equation}
|\psi(t)\rangle = c_1(t) |1\rangle + c_2(t) |2\rangle,
\end{equation}
where $|1\rangle$ and $|2\rangle$ form a basis (either diabatic or adiabatic), and $c_1(t)$ and $c_2(t)$ are complex coefficients dependent on time.

The expectation value of the Hamiltonian for this state is:
\begin{equation}
\langle \psi(t) | \hat{H}(\mathbf{R}) | \psi(t) \rangle = |c_1|^2 H_{11} + |c_2|^2 H_{22} + 2\text{Re}(c_1^* c_2 H_{12}),
\end{equation}
where $H_{ij} = \langle i | \hat{H}(\mathbf{R}) | j \rangle$ are the matrix elements of the Hamiltonian.

The forces driving the nuclear motion are then computed as:
\begin{align}
F_x &= -\frac{\partial}{\partial x} \langle \psi(t) | \hat{H}(\mathbf{R}) | \psi(t) \rangle, \\
F_y &= -\frac{\partial}{\partial y} \langle \psi(t) | \hat{H}(\mathbf{R}) | \psi(t) \rangle.
\end{align}

where $\alpha = a+1$ and $\beta = 4(1-e^2)$.

When nuclear motion follows a closed loop around a conical intersection, the adiabatic electronic wavefunction acquires a geometric phase. This phase is related to the Berry connection, defined as:
\begin{equation}
\mathbf{A}_\pm(\mathbf{R}) = i\langle \psi_\pm(\mathbf{R})|\nabla_\mathbf{R}|\psi_\pm(\mathbf{R})\rangle,
\end{equation}
and the Berry phase acquired along a closed path $C$ is:
\begin{equation}
\gamma_\pm = \oint_C \mathbf{A}_\pm(\mathbf{R}) \cdot d\mathbf{R}.
\end{equation}

For a conical intersection (where $E(x, y) = 0$), the Berry phase for a loop encircling the origin is $\gamma_\pm = \pi$, corresponding to a sign change in the electronic wavefunction. For other types of intersections, the Berry phase may be different and can depend on the path geometry.

\subsection{Quantum-Classical Lagrangian and Equations of Motion}

We present a rigorous derivation of the equations of motion for mixed quantum-classical dynamics, starting from first principles. Consider a system with quantum electronic state $|\psi(t)\rangle$ and classical nuclear coordinates $\mathbf{R}(t) = (x(t), y(t))$.

\subsubsection{The Fundamental Lagrangian}

We begin with the quantum-classical Lagrangian \cite{todorov2001time,halliday2022manifold}:
\begin{equation}
L = \langle \psi | i\hbar \frac{\partial}{\partial t} - \hat{H}(\mathbf{R}) | \psi \rangle + \frac{1}{2}M\dot{\mathbf{R}}^2,
\end{equation}
where $M$ is the nuclear mass and $\hat{H}(\mathbf{R})$ is the electronic Hamiltonian parameterized by nuclear coordinates.

To ensure the Lagrangian is real-valued, we employ the time-symmetrized form:
\begin{equation}
L = \frac{i\hbar}{2}\left(\langle \psi | \frac{\partial \psi}{\partial t} \rangle - \langle \frac{\partial \psi}{\partial t} | \psi \rangle\right) - \langle \psi | \hat{H}(\mathbf{R}) | \psi \rangle + \frac{1}{2}M\dot{\mathbf{R}}^2.
\end{equation}

The time-symmetrization is justified by the normalization constraint $\langle \psi | \psi \rangle = 1$, which gives:
\begin{equation}
\frac{d}{dt}\langle \psi | \psi \rangle = \langle \frac{\partial \psi}{\partial t} | \psi \rangle + \langle \psi | \frac{\partial \psi}{\partial t} \rangle = 0,
\end{equation}
confirming that $\langle \psi | \frac{\partial \psi}{\partial t} \rangle = -\langle \frac{\partial \psi}{\partial t} | \psi \rangle^*$ and the Lagrangian is purely real.

\subsubsection{Adiabatic Basis Expansion}

We expand the quantum state in the instantaneous eigenbasis of $\hat{H}(\mathbf{R})$:
\begin{equation}
|\psi(t)\rangle = \sum_n c_n(t) |n(\mathbf{R}(t))\rangle,
\end{equation}
where the adiabatic states satisfy:
\begin{equation}
\hat{H}(\mathbf{R})|n(\mathbf{R})\rangle = E_n(\mathbf{R})|n(\mathbf{R})\rangle.
\end{equation}

The time derivative of the state is:
\begin{equation}
\frac{\partial|\psi\rangle}{\partial t} = \sum_n \dot{c}_n |n\rangle + \sum_{n,\alpha} c_n \frac{\partial |n\rangle}{\partial R_\alpha}\dot{R}_\alpha,
\end{equation}
where we sum over coordinate components $\alpha \in \{x, y\}$.

\subsubsection{Expanding the Lagrangian Terms}

Computing $\langle \psi | \frac{\partial \psi}{\partial t} \rangle$ using orthonormality $\langle m|n\rangle = \delta_{mn}$:
\begin{equation}
\langle \psi | \frac{\partial \psi}{\partial t} \rangle = \sum_n c_n^* \dot{c}_n + \sum_{m,n,\alpha} c_m^* c_n \langle m | \frac{\partial n}{\partial R_\alpha} \rangle \dot{R}_\alpha.
\end{equation}

We define the non-adiabatic coupling matrix elements:
\begin{equation}
d_{mn}^\alpha = \langle m | \frac{\partial}{\partial R_\alpha} | n \rangle,
\end{equation}
and the Berry connection (for diagonal elements):
\begin{equation}
A_{n\alpha} = i\langle n | \frac{\partial}{\partial R_\alpha} | n \rangle.
\end{equation}

Note that $A_{n\alpha}$ is real because differentiating $\langle n | n \rangle = 1$ gives:
\begin{equation}
\langle \frac{\partial n}{\partial R_\alpha} | n \rangle + \langle n | \frac{\partial n}{\partial R_\alpha} \rangle = 0,
\end{equation}
implying $\langle n | \frac{\partial n}{\partial R_\alpha} \rangle$ is purely imaginary.

The time-symmetrized kinetic term becomes:

\begin{equation}
\begin{split}
    & \frac{i\hbar}{2}\left(\langle \psi | \frac{\partial \psi}{\partial t} \rangle - \langle \frac{\partial \psi}{\partial t} | \psi \rangle\right) = \\
    & \qquad i\hbar \sum_n c_n^* \dot{c}_n \\
    & \qquad + \hbar\sum_{n,\alpha} |c_n|^2 A_{n\alpha} \dot{R}_\alpha \\
    & \qquad + i\hbar \sum_{\substack{m,n \\ m \neq n}} \sum_\alpha c_m^* c_n d_{mn}^\alpha \dot{R}_\alpha.
\end{split}
\end{equation}

The Hamiltonian expectation value is:
\begin{equation}
\langle \psi | \hat{H} | \psi \rangle = \sum_n |c_n|^2 E_n(\mathbf{R}).
\end{equation}

Thus, the complete Lagrangian is:
\begin{equation}
\begin{split}
    L = & \, i\hbar \sum_n c_n^* \dot{c}_n + \hbar\sum_{n,\alpha} |c_n|^2 A_{n\alpha} \dot{R}_\alpha \\
    & + i\hbar \sum_{\substack{m,n \\ m \neq n}} \sum_\alpha c_m^* c_n d_{mn}^\alpha \dot{R}_\alpha \\
    & - \sum_n |c_n|^2 E_n + \frac{1}{2}M\dot{\mathbf{R}}^2.
\end{split}
\end{equation}

\subsection{Euler-Lagrange Equations}

\subsubsection{Quantum Evolution}

The Euler-Lagrange equation for $c_n^*$ (treating $c_n$ and $c_n^*$ as independent variables):
\begin{equation}
\frac{d}{dt}\frac{\partial L}{\partial \dot{c}_n^*} - \frac{\partial L}{\partial c_n^*} = 0.
\end{equation}

Computing the derivatives:
\begin{equation}
\frac{\partial L}{\partial \dot{c}_n^*} = i\hbar c_n, \quad \frac{d}{dt}(i\hbar c_n) = i\hbar \dot{c}_n,
\end{equation}

\begin{equation}
\frac{\partial L}{\partial c_n^*} = i\hbar \dot{c}_n + \hbar\sum_\alpha c_n A_{n\alpha}\dot{R}_\alpha + i\hbar \sum_{m \neq n} \sum_\alpha c_m d_{nm}^\alpha \dot{R}_\alpha - c_n E_n.
\end{equation}

This yields the time-dependent Schrödinger equation in the adiabatic basis:
\begin{equation}
i\hbar \dot{c}_n = E_n c_n - \hbar\sum_\alpha A_{n\alpha}\dot{R}_\alpha c_n - i\hbar \sum_{m \neq n} \sum_\alpha d_{nm}^\alpha \dot{R}_\alpha c_m.
\end{equation}

\subsubsection{Classical Evolution}

For the nuclear coordinates, the Euler-Lagrange equation is:
\begin{equation}
\frac{d}{dt}\frac{\partial L}{\partial \dot{R}_\alpha} - \frac{\partial L}{\partial R_\alpha} = 0.
\end{equation}

Computing $\frac{\partial L}{\partial \dot{R}_\alpha}$:
\begin{equation}
\frac{\partial L}{\partial \dot{R}_\alpha} = \hbar\sum_n |c_n|^2 A_{n\alpha} + i\hbar \sum_{\substack{m,n \\ m \neq n}} c_m^* c_n d_{mn}^\alpha + M\dot{R}_\alpha.
\end{equation}

Taking the time derivative:

\begin{equation}
\begin{split}
    \frac{d}{dt}\frac{\partial L}{\partial \dot{R}_\alpha} = & \hbar\sum_n \frac{d}{dt}(|c_n|^2) A_{n\alpha} \\
    & + \hbar\sum_{n,\beta} |c_n|^2 \frac{\partial A_{n\alpha}}{\partial R_\beta}\dot{R}_\beta + \text{NAC terms} + M\ddot{R}_\alpha,
\end{split}
\end{equation}

where $\frac{d}{dt}(|c_n|^2) = \dot{c}_n^* c_n + c_n^* \dot{c}_n$.

Computing $\frac{\partial L}{\partial R_\alpha}$ involves derivatives of all $\mathbf{R}$-dependent terms:

\begin{equation}
\begin{split}
    \frac{\partial L}{\partial R_\alpha} = & -\sum_n |c_n|^2 \frac{\partial E_n}{\partial R_\alpha} \\
    & + \hbar\sum_{n,\beta} |c_n|^2 \frac{\partial A_{n\beta}}{\partial R_\alpha}\dot{R}_\beta + \text{NAC derivative terms}.
\end{split}
\end{equation}


After substantial algebra (using the quantum equation of motion to eliminate $\dot{c}_n$ terms), we obtain:
\begin{equation}
M\ddot{R}_\alpha = -\sum_n |c_n|^2 \frac{\partial E_n}{\partial R_\alpha} + \sum_{n,\beta} |c_n|^2 \hbar\Omega_{n,\alpha\beta}\dot{R}_\beta + F_\alpha^{\text{NAC}},
\end{equation}
where the Berry curvature is:
\begin{equation}
\Omega_{n,\alpha\beta} = \frac{\partial A_{n\beta}}{\partial R_\alpha} - \frac{\partial A_{n\alpha}}{\partial R_\beta},
\end{equation}
and the force due to non-adiabatic couplings is:
\begin{equation}
F_\alpha^{\text{NAC}} = 2\hbar\sum_{m \neq n} \text{Re}[c_m^* c_n (E_n - E_m) d_{mn}^\alpha].
\end{equation}

\subsection{Final Equations of Motion}

The complete set of coupled quantum-classical equations is:

\subsubsection{Classical Motion}

\begin{equation}
\begin{split}
    M\ddot{R}_\alpha = & -\sum_n |c_n|^2 \frac{\partial E_n}{\partial R_\alpha} + \sum_{n,\beta} |c_n|^2 \hbar\Omega_{n,\alpha\beta}\dot{R}_\beta \\
    & + 2\hbar\sum_{m \neq n} \text{Re}[c_m^* c_n (E_n - E_m) d_{mn}^\alpha].
\end{split}
\end{equation}


\subsubsection{Quantum Evolution}
\begin{equation}
i\hbar \dot{c}_n = E_n c_n - \hbar\sum_\alpha A_{n\alpha}\dot{R}_\alpha c_n - i\hbar \sum_{m \neq n, \alpha} d_{nm}^\alpha \dot{R}_\alpha c_m.
\end{equation}

\subsection{Physical Interpretation of Force Terms}

\subsubsection{Ehrenfest Force}
The first term represents the standard Ehrenfest force:
\begin{equation}
\mathbf{F}^{\text{Ehr}} = -\sum_n |c_n|^2 \nabla E_n(\mathbf{R}),
\end{equation}
which is the expectation value of the force operator in the adiabatic basis.

\subsubsection{Berry Curvature Force}
The second term is the force due to Berry curvature:
\begin{equation}
\mathbf{F}^{\text{Berry}} = \sum_n |c_n|^2 \hbar (\mathbf{v} \times \mathbf{B}_n),
\end{equation}
where $\mathbf{B}_n$ is the Berry curvature vector. In our 2D case:
\begin{equation}
F_x^{\text{Berry}} = \sum_n |c_n|^2 \hbar \Omega_{n,xy} v_y, \quad F_y^{\text{Berry}} = -\sum_n |c_n|^2 \hbar \Omega_{n,xy} v_x.
\end{equation}

\subsubsection{Non-Adiabatic Coupling Force}
The third term arises from transitions between adiabatic states and is proportional to the electronic coherence $\text{Re}(c_m^* c_n)$ and the energy gap $(E_n - E_m)$.

\textbf{Important Note:} There is no ``geometric force'' proportional to $\nabla \cdot \mathbf{A}$. Such terms are gauge-dependent artifacts that cancel exactly in a correct derivation. Only the gauge-invariant Berry curvature appears in the physical equations of motion.

\subsection{Berry Curvature and Force Corrections}
\subsection{Berry Phase}

\begin{figure}[ht]
    \centering
    \includegraphics[width=0.45\textwidth]{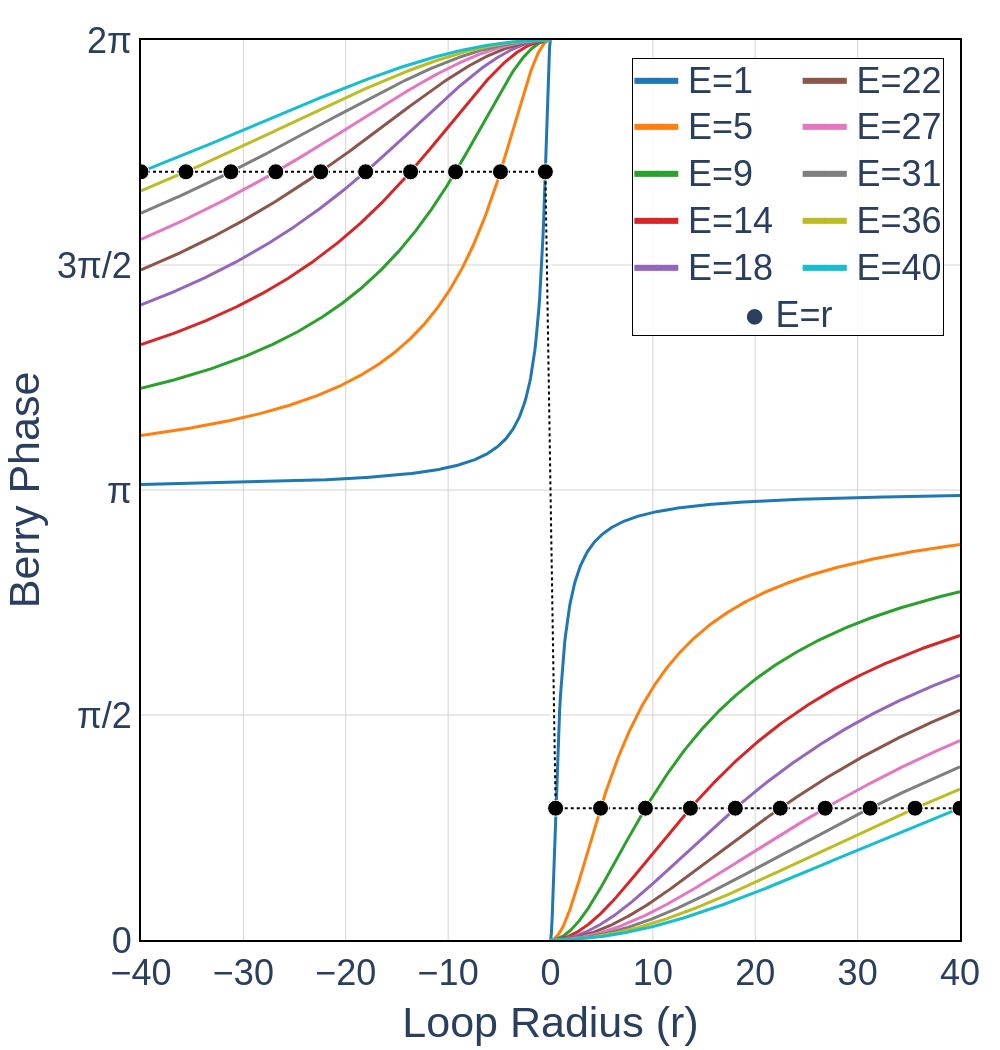}
    \caption{Berry phase variation with loop radius for different energy gaps $E$ in our model $2X2$ Hamiltonian, showing characteristic behavior around the avoided crossings. Smaller $E$ values produce sharper transitions in the phase, while larger $E$ values lead to more gradual changes, reflecting the topographical nature of the phase accumulation in quantum systems. The dots represents the Berry's phase for the avoided crossings at the points $E=r$, revealing the phase due to elliptical intersection.}
    \label{fig:berry_phase}
\end{figure}

The Berry phase characterizes the geometric phase acquired by quantum eigenstates under adiabatic evolution around a closed loop in parameter space. Figure~\ref{fig:berry_phase} illustrates how the Berry phase varies with loop radius for different energy gaps ($E$ values) between diabatic states, demonstrating the transition from topological to geometric behavior as one moves away from the conical intersection.

The Berry phase for a closed loop around the $z$-axis can be calculated as:
\begin{align}\label{eq:berry phase}
\gamma_- &= \pi\left(1 - \frac{E\alpha}{\sqrt{E^2\alpha^2 + r^2\beta}}\right) \\
\gamma_+ &= \pi\left(1 + \frac{E\alpha}{\sqrt{E^2\alpha^2 + r^2\beta}}\right)
\end{align}
where $\gamma_-$ corresponds to the lower cone and $\gamma_+$ to the upper cone and the phase accumulation occurred in the clockwise direction, if we integrated counter clockwise, the phase accumulation would observe a sign change.

It is well known in the literature that the geometric phase can be tuned and that it can not always be a topological invariant like in the case of gapped graphene where the geometric phase also depends on the tunable band gap \cite{de2010intersections,urru2015tunability}. For the case of avoided crossings as in \ref{fig:berry_phase}, the geometric phase is dependent on the background energy landscape and the path taken to accumulate what is known in literature as a marginal Berry phase \cite{oh2008entanglement}. The parameters in the Hamiltonian provide control over the geometric phase and thus the surface topology, allowing the Hamiltonian to be tuned to a path invariant phase that is non-trivial and other than $\pi$. The topological phase of the elliptic intersection manifests itself due to an exact balance between the magnitude of the diabatic couplings and the diagonal terms in the hamiltonian or the energy levels $(E)$.

From the expressions for the berry phases we can quickly identify that for the case of elliptic intersection, that is, when $E=r$, the system accumulates a path-invariant Berry phase with the value $\pi(1-\frac{\alpha}{\sqrt{\alpha^2+\beta}})$. The Berry phase (\ref{eq:berry phase}) is tunable due to the parameters $a$ and $e$ and the dynamical implication of this is the tunability of the gauge-invariant Berry curvature and hence the force. Although presented as distinct categories, a elliptic intersection can be realized as a special case within the avoided crossing regime, when $E=r$. It is interesting to note that within the avoided crossing regime there exist trajectories $(E=r)$ that are topological and not only geometric. The avoided crossings are responsible for the marginal berry phases \cite{oh2008entanglement}, which are neither topological nor energy dependent; we establish topological phases due to elliptic intersection or $E=r$ within the avoided crossing regime \ref{fig:berry_phase}.

The conventional Ehrenfest dynamics captures the mean-field force arising from the electronic potential energy surfaces; it does not account for the topological effects of the geometric phase directly and accurately. To incorporate these effects, we introduce a correction to the forces based on the Berry curvature. The Berry curvature is defined as the curl of the Berry connection:
\begin{equation}
\mathbf{B}{_\pm}(\mathbf{R}) = \nabla_\mathbf{R} \times \mathbf{A}_\pm(\mathbf{R}),
\end{equation}
or, in our two-dimensional case:
\begin{equation}
B_{z\pm} = \frac{\partial A_{\pm,y}}{\partial x} - \frac{\partial A_{\pm,x}}{\partial y}.
\end{equation}

where $\alpha = a+1$ and $\beta = 4(1-e^2)$.
The Berry curvature gives rise to an additional force term, analogous to the Lorentz force experienced by a charged particle in a magnetic field:
\begin{equation}
\mathbf{F}_{\text{Berry}} = \mathbf{v} \times \mathbf{B}_\pm(\mathbf{R}),
\end{equation}
where $\mathbf{v} = \dot{\mathbf{R}}$ is the nuclear velocity.

In our two-dimensional case, this Berry force takes the form:
\begin{align}
F_{x,\text{Berry}} &= v_y B_{z\pm}, \\
F_{y,\text{Berry}} &= -v_x B_{z\pm}.
\end{align}

Incorporating this correction, the complete equations of motion for the nuclei become:
\begin{align}
M\ddot{x} &= -\frac{\partial}{\partial x} \langle \psi(t) | \hat{H}(\mathbf{R}) | \psi(t) \rangle + v_y B_\pm, \\
M\ddot{y} &= -\frac{\partial}{\partial y} \langle \psi(t) | \hat{H}(\mathbf{R}) | \psi(t) \rangle - v_x B_\pm.
\end{align}

This Berry force correction ensures that the nuclear trajectories properly account for the topological effects of the geometric phase, leading to more accurate dynamics, especially in regions where the Berry curvature is significant.
\subsection{Pre-looping Trajectory Initialization}

\begin{figure}[ht]
\centering
\includegraphics[width=0.5\textwidth]{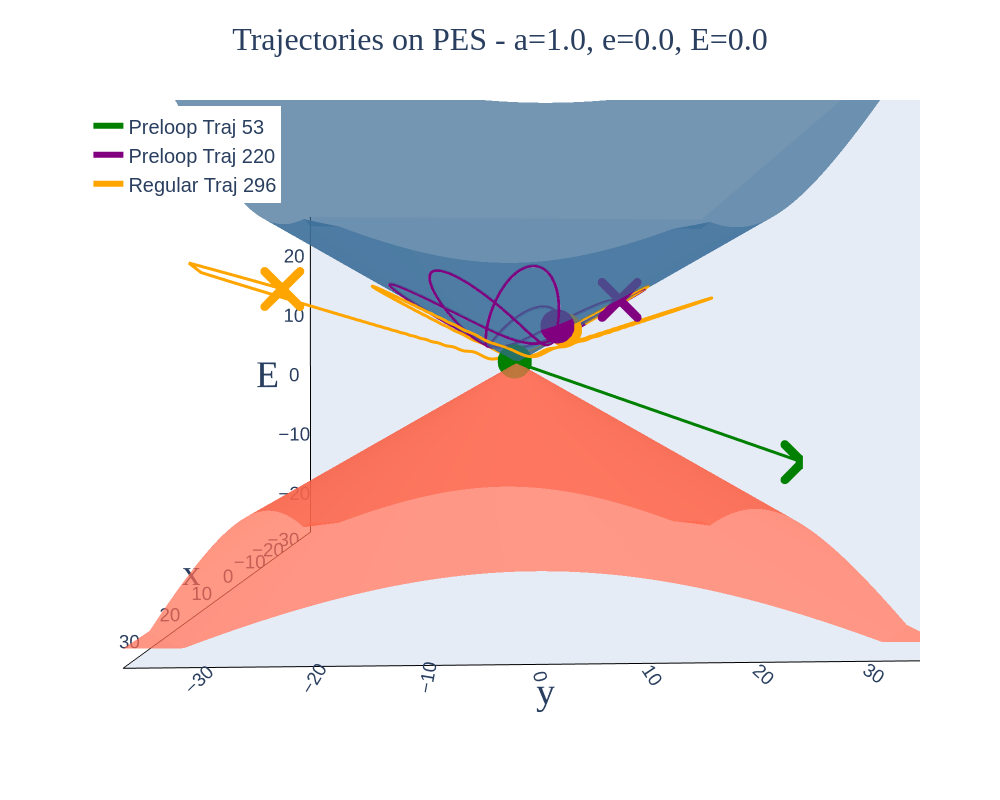}
\caption{Trajectories evolving over simulation time as Gaussian wavepackets in the adiabatic basis. The trajectories are plotted with the adiabatic potential energy surfaces in represents the background energy landscape. The visualization shows three distinct trajectories (green, purple, and yellow) with their respective starting $(\bullet)$ and ending $(\times)$ points marked. The blue and red surfaces represent the upper and lower adiabatic energy surfaces, respectively. Simulation performed with 500 total trajectories - 250 regular trajectories and 250 are pre-looping.}
\label{fig:prelooped}
\end{figure}

In our computational framework, we introduce the \textbf{pre-looping trajectory initialization scheme}. This approach is designed to efficiently initialize a trajectory with the berry's phase adiabatically accumulated as a result of evolving on a closed loop over the potential energy surface. Traditional initialization methods often result in trajectories that either miss the crossing region or traverse it without capturing the full topological effect. The pre-looping scheme addresses this limitation by applying the analytical Berry phase directly to the wavefunction at the simulation's outset ($t=0$).

This initialization highlights the topological properties of different types of crossings. For a conical intersection ($E=0$), the Berry curvature vanishes in our two-dimensional model and fixes the associated Berry phase to $\pi$. Our framework, however, also accommodates elliptic intersections ($E=r$), a class of degeneracy characterized by a non-vanishing and parametrically tunable Berry curvature. This feature allows for the study of systems that accumulate a path-invariant Berry phase with a value other than $\pi$ in 2 dimensions.

\subsubsection{The Pre-looping Concept}
The significance of the geometric phase in quantum dynamics extends beyond the technical challenge of resolving the double-valued electronic wave function, a problem often addressed by employing a resolution of the identity to restore a single-valued representation \cite{mead1979determination, Ryabinkin2017Geometric}. More fundamentally, a failure to account for the phase consistency imposed by the geometric phase leads to demonstrably inconsistent population dynamics \cite{akimov2018simple}. The fundamental idea behind pre-looping is to initialize the quantum state with the geometric phase that would be acquired during a complete adiabatic loop around the crossing point or degeneracy. The pre-looping initialization scheme tries to imply that the adiabatically accumulated history can be used as the initialization unlike the ab-initio schemes where the particles have no past experience. The topological memory is encoded as the initial phase that is gauge invariant and doesn't explicitly affect the dynamics but, represents the initial orientation when the dynamics begin and thus reflects as change in the population dynamics.

The nuclear coordinates and momenta are carefully selected to place the trajectory in a region where these topological effects are significant. The initialization is performed in several steps:

\begin{enumerate}
    \item Select an initial radius $r$ from the crossing point. This radius must be large enough to avoid the singularity but small enough for the system to be influenced by the strong nonadiabatic coupling near the crossing.
    \item Choose an angular position $\theta$ on the circle (or ellipse, for $e \neq 0$) of radius $r$.
    \item Calculate the position coordinates $(x_0, y_0)$ based on $r$ and $\theta$, accounting for the eccentricity parameter $e$:
    \begin{align}
        x_0 &= r\cos\theta, \label{eq:x0_final} \\
        y_0 &= \frac{r\sin\theta}{\sqrt{1-e^2}}. \label{eq:y0_final}
    \end{align}
    \item Compute the Berry curvature at this position to determine the strength of the topological effect.
    \item Set the initial momentum perpendicular to the radial vector, with its direction determined by the sign of the Berry curvature:
    \begin{align}
        p_{x0} &= -\text{sign}(B_\pm) v_{\text{magnitude}} \sin\theta, \label{eq:px0_final} \\
        p_{y0} &= \text{sign}(B_\pm) v_{\text{magnitude}} \cos\theta \sqrt{1-e^2}, \label{eq:py0_final}
    \end{align}
    where $v_{\text{magnitude}}$ is an adjustable velocity scale.
    \item Initialize the electronic state in either the lower or upper adiabatic state.
    \item Apply a phase factor to the initial electronic state to simulate post-looping dynamics:
    \begin{equation}
        |\psi(0)\rangle = e^{i\gamma_\pm} |\psi_\pm(\mathbf{R}_0)\rangle, \label{eq:psi0_final}
    \end{equation}
    where $\gamma_\pm$ is the pre-calculated, analytical Berry phase for the chosen state and loop radius.
\end{enumerate}

The pre-looping approach provides a controlled method for studying the influence of the geometric phase on nonadiabatic dynamics. This initialization ensures the trajectory begins with the correct phase accumulated for forming a loop of constant-energy in the X-Y plane. This setup imparts an initial velocity that biases the trajectory toward an orbital path, driven by the balance between the potential gradient and the Berry force, which acts as a local, effective magnetic field.

\section{Computational Implementation}
\subsection{Numerical Integration Scheme}

Our computational framework implements the modified Ehrenfest dynamics with Berry curvature corrections and non-adiabatic coupling terms using robust numerical integration techniques. For the nuclear equations of motion, we employ a modified velocity Verlet algorithm that preserves energy conservation while properly accounting for the Berry force:

\begin{align}
\mathbf{v}(t + \frac{\Delta t}{2}) &= \mathbf{v}(t) + \frac{\Delta t}{2M} \mathbf{F}(t), \\
\mathbf{R}(t + \Delta t) &= \mathbf{R}(t) + \Delta t \mathbf{v}(t + \frac{\Delta t}{2}), \\
\mathbf{F}(t + \Delta t) &= \mathbf{F}_{\text{Ehrenfest}}(t + \Delta t) + \mathbf{F}_{\text{Berry}}(t + \Delta t), \\
\mathbf{v}(t + \Delta t) &= \mathbf{v}(t + \frac{\Delta t}{2}) + \frac{\Delta t}{2M} \mathbf{F}(t + \Delta t),
\end{align}

where $\mathbf{F}_{\text{Ehrenfest}}$ is the standard Ehrenfest force derived from the gradient of the electronic Hamiltonian's expectation value, and $\mathbf{F}_{\text{Berry}}$ is the force correction based on the Berry curvature.

For the electronic wavefunction, we use a fourth-order Runge-Kutta (RK4) method to solve the time-dependent Schrödinger equation with non-adiabatic coupling terms:


To ensure numerical stability and accuracy, especially near conical intersections where both the potential energy surfaces and non-adiabatic couplings can vary rapidly, we implement adaptive time-stepping. The time step is adjusted based on the energy conservation error:

\begin{equation}
\Delta t_{\text{new}} = \Delta t \times \min\left(2, \max\left(0.1, \sqrt{\frac{\epsilon_{\text{target}}}{\epsilon_{\text{actual}}}}\right)\right),
\end{equation}

where $\epsilon_{\text{target}}$ is the target energy conservation error and $\epsilon_{\text{actual}}$ is the actual error in the current step. This approach ensures that the time step is reduced in regions of rapid variation while allowing larger steps in smoother regions.

\subsection*{Berry Curvature Calculation}

A critical aspect of our simulation framework is the accurate computation of the Berry curvature, which is fundamental to the geometric phase force. To this end, we employ a hybrid numerico-analytical scheme wherein the Berry curvature is determined from a direct analytical expression, while the first-order non-adiabatic couplings are computed numerically. This strategy is adopted to mitigate the numerical instabilities inherent in calculating the second-order spatial derivatives that define the curvature, thereby ensuring a robust and precise evaluation of the geometric phase effects.

The analytical expression for the Berry curvature is derived \textit{a priori} using a symbolic algebra system. The derivation proceeds via the following steps:

\begin{enumerate}
    \item \textbf{Analytical Eigendecomposition}: The two-level electronic Hamiltonian, $\hat{H}(\mathbf{R})$, is analytically diagonalized to obtain closed-form expressions for its adiabatic eigenvectors, $\ket{\psi_\pm(\mathbf{R})}$, and eigenvalues as functions of the nuclear coordinates $\mathbf{R} = (x, y)$.

    \item \textbf{Berry Connection}: The components of the Berry connection vector, $\mathbf{A}_\pm(\mathbf{R})$, are subsequently derived by analytically differentiating the eigenvectors according to the formal definition:
    \begin{equation}
    \mathbf{A}_\pm(\mathbf{R}) = -i \bra{\psi_\pm(\mathbf{R}) }| \nabla_\mathbf{R} |\ket{ \psi_\pm(\mathbf{R})}
    \end{equation}

    \item \textbf{Berry Curvature}: Finally, the Berry curvature is obtained by analytically computing the curl of the Berry connection vector, $\mathbf{\Omega}_\pm(\mathbf{R}) = \nabla_\mathbf{R} \times \mathbf{A}_\pm(\mathbf{R})$. For the two-dimensional parameter space of the nuclei, the only non-vanishing component is perpendicular to the plane of motion, yielding the scalar curvature:
    \begin{equation}
    B_{z, \pm}(\mathbf{R}) = \frac{\partial A_{\pm,y}}{\partial x} - \frac{\partial A_{\pm,x}}{\partial y}
    \end{equation}
\end{enumerate}

The resulting algebraic expression for $B_{z, \pm}(\mathbf{R})$, though complex, is exact. This function is then implemented directly within the simulation code, allowing for its efficient evaluation at any point in the nuclear configuration space. This analytical treatment of the Berry curvature stands in contrast to the numerical evaluation of the first-order non-adiabatic couplings, which are less susceptible to numerical error. This hybrid approach provides an optimal balance of computational accuracy and stability for simulating the non-adiabatic quantum dynamics.

\subsection*{Non-Adiabatic Coupling Terms}

In addition to forces arising from the Berry curvature, our simulation framework also incorporates non-adiabatic coupling terms (NACTs) to accurately model the dynamics in regions where the Born-Oppenheimer approximation breaks down. These terms are essential for describing the transitions between different electronic states induced by nuclear motion.

The first-order non-adiabatic coupling vector between two adiabatic electronic states, $\ket{\psi_i(\mathbf{R})}$ and $\ket{\psi_j(\mathbf{R})}$, is formally defined as $\mathbf{d}_{ij}(\mathbf{R}) = \bra{\psi_i(\mathbf{R}) }| \nabla_\mathbf{R} |\ket{ \psi_j(\mathbf{R})}$. In our implementation, we compute this vector using a more numerically stable formula derived from the Hellmann-Feynman theorem:
\begin{equation}
\mathbf{d}_{ij}(\mathbf{R}) = \frac{\bra{\psi_i(\mathbf{R}) }| \nabla_\mathbf{R} \hat{H}(\mathbf{R}) |\ket{ \psi_j(\mathbf{R})}}{E_j(\mathbf{R}) - E_i(\mathbf{R})}, \quad \text{for } i \neq j
\end{equation}
Here, $\hat{H}(\mathbf{R})$ is the electronic Hamiltonian at a given nuclear geometry $\mathbf{R}$, while $E_i$ and $E_j$ are the corresponding adiabatic energies. This approach avoids the direct numerical differentiation of the eigenvectors. For our two-level system, the key coupling is between the lower ($-$) and upper ($+$) states, giving $\mathbf{d}_{-+}(\mathbf{R})$. The required gradients of the Hamiltonian, $\nabla_\mathbf{R} \hat{H}$, are calculated using a central difference method.

These coupling terms are incorporated into the simulation in two critical ways:
\begin{enumerate}
    \item \textbf{Electronic State Evolution:} The NACTs govern the time evolution of the electronic wavefunction $\ket{\psi(t)}$. The nuclear velocity vector $\dot{\mathbf{R}}$ couples with the NACT vector $\mathbf{d}_{ij}$ to drive population transfer between the electronic states. This is included in the time-dependent Schrödinger equation:
    \begin{equation}
    i\hbar \frac{d}{dt} \ket{\psi(t)} = \hat{H}(\mathbf{R}(t)) \ket{\psi(t)} - i\hbar \sum_{i \neq j} c_j(t) (\dot{\mathbf{R}} \cdot \mathbf{d}_{ij}(\mathbf{R})) \ket{\psi_i(\mathbf{R})}
    \end{equation}
    where $c_j(t)$ are the expansion coefficients of $\ket{\psi(t)}$ in the adiabatic basis.

    \item \textbf{Nuclear Forces:} The NACTs also contribute a force component acting on the nuclei, which is particularly significant when the system is in a superposition of states. As implemented in the code, this force is given by:
    \begin{equation}
    \mathbf{F}_{\text{NAC}} = -2 \text{Re}(c_-^* c_+) \mathbf{d}_{-+}(\mathbf{R})
    \end{equation}
    where $\text{Re}(c_-^* c_+)$ represents the real part of the electronic coherence between the lower and upper adiabatic states. This force component is added to the standard Ehrenfest force and other geometric phase corrections.
\end{enumerate}

The inclusion of these first-order NACTs enables the framework to capture the dynamics of electronic transitions with high fidelity, especially near conical intersections where such effects are dominant. This provides a more complete physical description of the system's evolution in non-adiabatic regimes.
\subsection{Ensemble Averaging and Observables}
To obtain statistically meaningful results, especially for systems with quantum interference effects, we employ ensemble averaging over multiple trajectories. Starting from the same initial electronic state but with different initial nuclear positions and momenta, we propagate an ensemble of trajectories and compute various observables:

\begin{enumerate}
    \item \textbf{Electronic Population Dynamics}: The population of the adiabatic states $|\psi_\pm(\mathbf{R})\rangle$ is calculated as:
    \begin{equation}
    P_\pm(t) = |\langle \psi_\pm(\mathbf{R}(t)) | \psi(t) \rangle|^2,
    \end{equation}
    which measures the probability of finding the system in each adiabatic state at time $t$.
    
    
    \item \textbf{Berry Curvature Distribution}: In Figure~\ref{fig:force} we visualize the Berry curvature as a function of nuclear coordinates to identify regions of strong topological effects. 
\end{enumerate}

\begin{figure*}[ht]
\centering
\includegraphics[width=\textwidth]{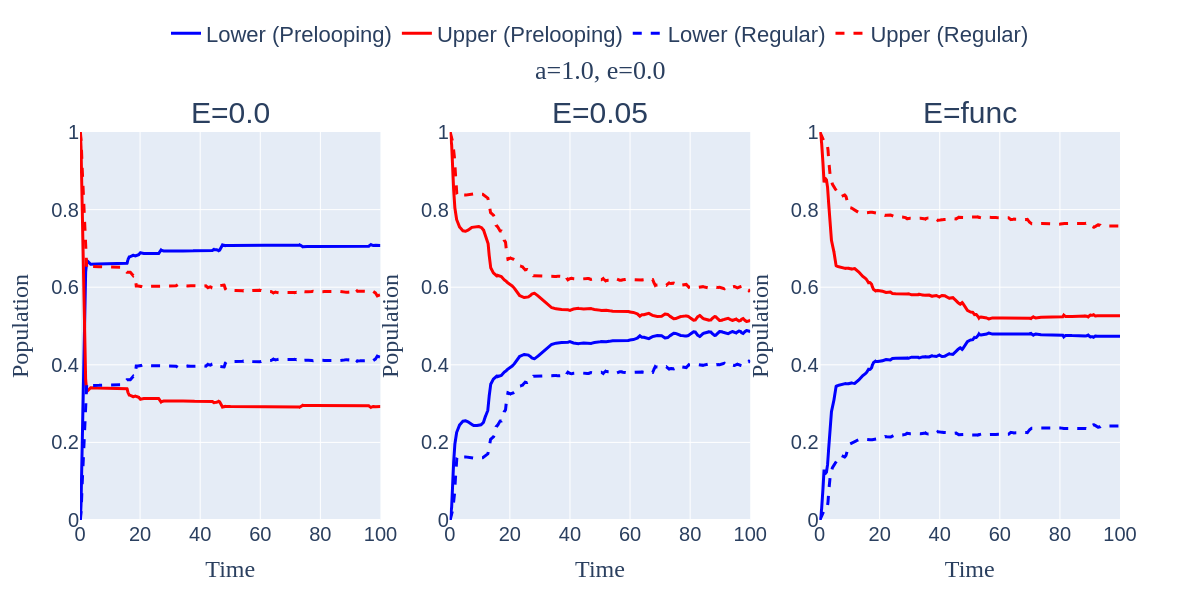}
\caption{ The panels represent: (a) conical intersection (E=0.0), (b) avoided crossing (E=0.05), and (c) elliptic intersection (E=func). The parameters $a=s=1$ and $e=0$ for all the three crossing types. Each panel shows the time evolution of adiabatic state populations, with solid lines representing prelooping trajectories and dashed lines representing regular trajectories. Blue lines correspond to the lower adiabatic state and red lines to the upper adiabatic state.}
\label{fig:population_dynamics}
\end{figure*}
\subsubsection{Advantages of Pre-looping}

\begin{itemize}
    \item \textbf{Efficient Sampling}: By incorporating the initial phase as the berry's phase for pre-looping, the method efficiently samples the geometric phase effect without requiring a large number of random initializations.
    \item \textbf{Control Over Crossing Distance}: The approach allows for precise control over how closely the trajectories approach the crossing point, allowing systematic studies of nonadiabatic effects as a function of proximity to the crossing.
    \item \textbf{Adaptability}: The scheme can be easily adapted to different types of crossings (conical, elliptic, avoided) by adjusting the initialization parameters.
\end{itemize}
\section{Results and Discussion}

\subsection{Force Analysis and Geometric Effects}
\begin{figure*}[ht]
    \centering
    \includegraphics[width=\textwidth]{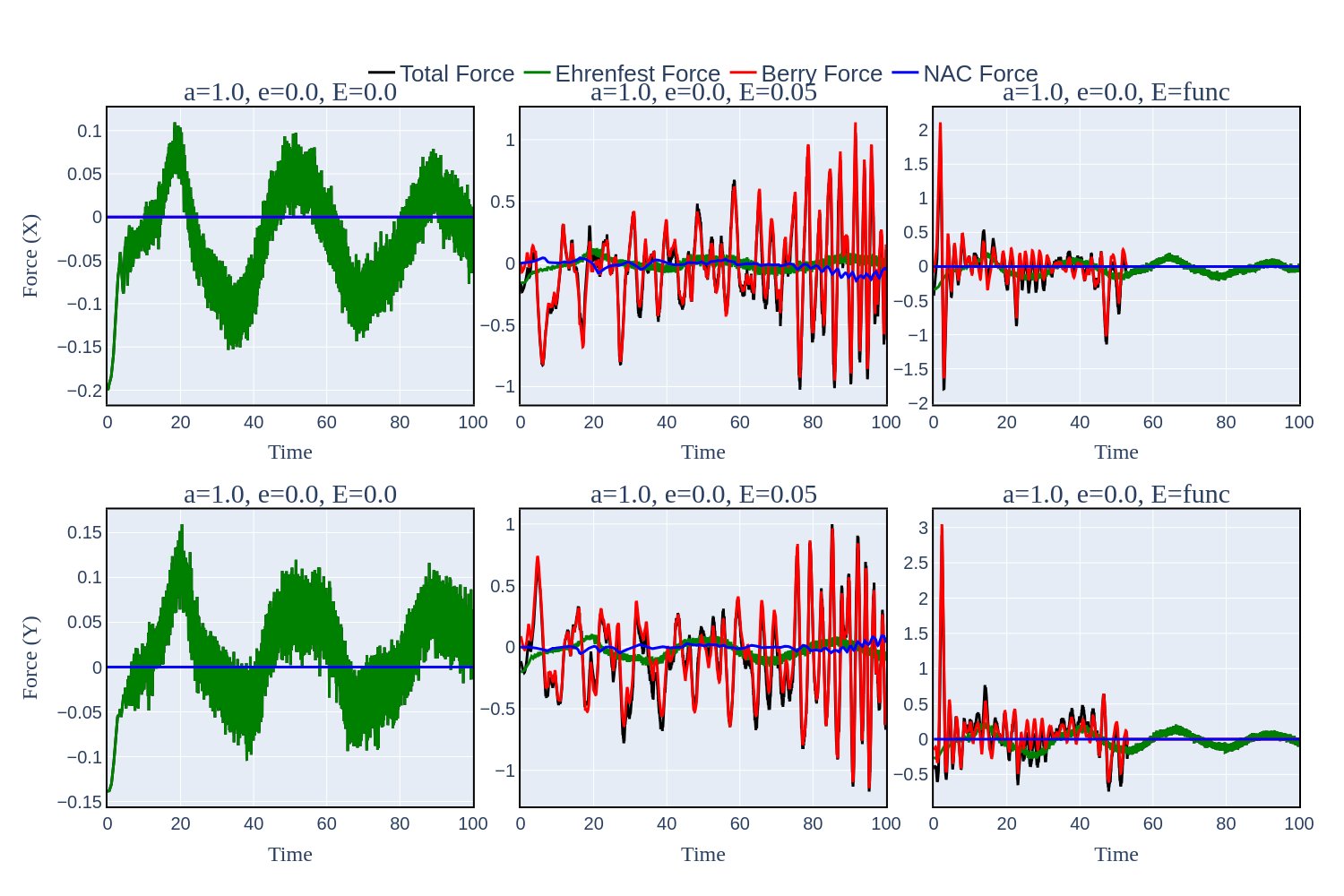}
    \caption{A comparative analysis of force components over time for three distinct non-adiabatic coupling regimes. The top row (X-components) and bottom row (Y-components) display the Total Force (black), Ehrenfest Force (green), Berry Force (red), and NAC Force (blue). The columns correspond to: (a) a conical intersection ($a=1.0, e=0.0, E=0.0$), where the Berry force is negligible; (b) an avoided crossing ($a=1.0, e=0.0, E=0.05$), where the Berry force becomes a dominant oscillatory component; and (c) a elliptic intersection ($a=1.0, e=0.0, E=\text{func}$), a degenerate crossing that supports a significant Berry force.}
    \label{fig:force_components}
\end{figure*}
The interplay between the classical (Ehrenfest) and quantum-geometric (Berry and NAC) forces is pivotal in determining the nuclear dynamics through regions of non-adiabatic coupling. Figure~\ref{fig:force_components} presents a comparative analysis of the force components for trajectories evolving through three archetypal crossing scenarios: a conical intersection, an avoided crossing, and a elliptic intersection.

For the \textbf{conical intersection} (Fig.~\ref{fig:force_components}a), the dynamics are overwhelmingly governed by the Ehrenfest forces. Notably, the Berry force is effectively zero throughout the simulation. This is a direct consequence of the topology of a standard 2D conical intersection, whose associated Berry curvature vanishes everywhere except for a singularity at the exact point of degeneracy.

In contrast, the \textbf{avoided crossing} (Fig.~\ref{fig:force_components}b) demonstrates a fundamentally different force landscape. The introduction of a finite energy gap ($z=0.05$) lifts the degeneracy and "activates" a non-zero Berry curvature in the surrounding region of the parameter space. Consequently, the Berry force emerges as a significant, oscillatory contributor to the total force, often opposing the Ehrenfest component and highlighting a scenario where geometric phase effects are crucial for accurately describing the nuclear dynamics.

The \textbf{elliptic intersection} (Fig.~\ref{fig:force_components}c) represents a topologically distinct class of degeneracy. Unlike the conical intersection, it permits a non-zero Berry curvature in a 2D parameter space while retaining a point of exact electronic degeneracy. This is clearly reflected in the dynamics, where the Berry force is a significant contributor from the outset. This case illustrates a physically rich scenario where both singular coupling effects (strong NAC forces) and non-trivial geometric phase effects (strong Berry forces) coexist and dictate the trajectory's evolution.

\subsection{Validation of Pre-looping Initialization Scheme}

To assess the efficacy of the pre-looping initialization scheme in probing regions of strong non-adiabatic coupling, we analyzed the time evolution of the adiabatic state populations. Figure~\ref{fig:population_dynamics} compares the population dynamics originating from pre-looped (solid lines) and regular (dashed lines) trajectories for the three crossing types. The regular trajectories represent the standard Ehrenfest method that randomly initializes position, momenta, and phase.

 In the conical intersection case (Fig.~\ref{fig:population_dynamics}a), pre-looped trajectories exhibit pronounced oscillations and achieve a much higher degree of population mixing compared to their regular counterparts. This demonstrates that by initializing trajectories with orbital momentum in regions of strong coupling, the system more effectively navigates the non-adiabatic region, enhancing transition probabilities.

This trend persists for both the avoided crossing (Fig.~\ref{fig:population_dynamics}b) and the elliptic intersection (Fig.~\ref{fig:population_dynamics}c). Although the dynamics are qualitatively different because of the distinct potential energy surfaces, the pre-looping scheme consistently results in greater final population transfer. This robustly validates our hypothesis that this initialization method preferentially samples the regions of configuration space where non-adiabatic couplings and geometric phase effects are most pronounced, thereby providing a more comprehensive description of the system's quantum dynamics.



\subsection{Analysis of Elliptic Intersections}

\begin{figure*}[ht]
\centering
\includegraphics[width=\textwidth]{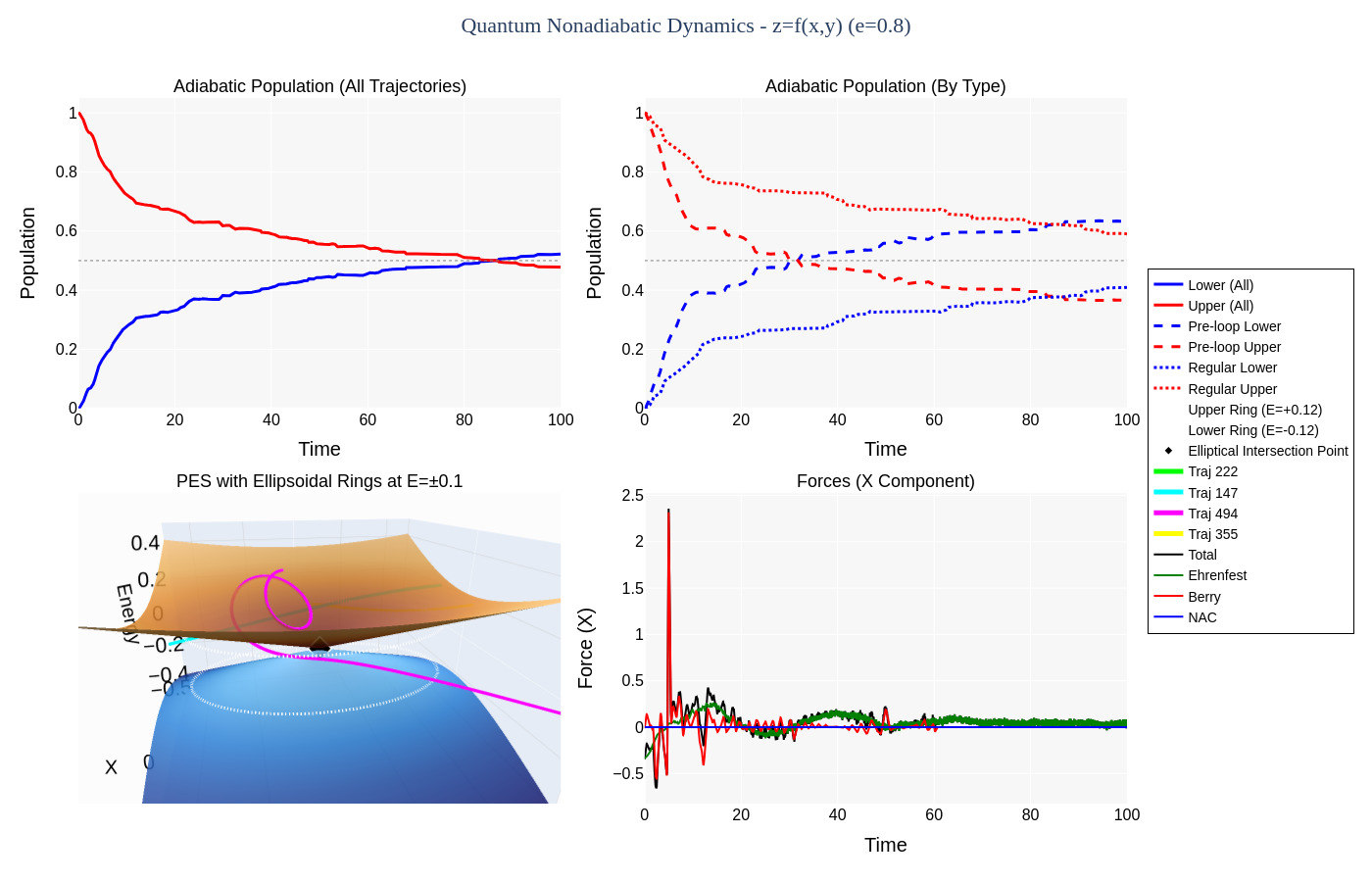}
\caption{Comprehensive analysis of nonadiabatic quantum dynamics at elliptic intersections [E=f(x,y)]. The 2×2 panel displays: (top left) adiabatic population dynamics averaged across all trajectories, showing gradual transfer from upper (red) to lower (blue) states; (top right) population comparison between pre-looped trajectories (solid lines) and regular trajectories (dashed lines), revealing enhanced population transfer in pre-looped simulations; (bottom left) three selected trajectories (magenta, cyan, yellow) visualized on potential energy surfaces, demonstrating distinct pathways through the coupling region; and (bottom right) comparison of quantum force components (Ehrenfest in green, Berry in red), highlighting the initial Berry force spike characteristic of geometric phase effects. System parameters: e=0.8, a=1.0, s=1.0, with 500 total trajectories (250 pre-looped, 250 regular) simulated over 20,000 time steps (dt=0.005).}
\label{fig:elliptic_comprehensive}
\end{figure*}

Figure~\ref{fig:elliptic_comprehensive} presents the elliptic intersection, defined by $E(x,y) = \sqrt{x^2 + (1-e^2)y^2}$, creates a crossing seam along a elliptic curve rather than a single point, leading to fundamentally different nonadiabatic dynamics compared to conical intersections.

The population dynamics (top panels) reveal characteristic features of elliptic intersections. The overall population evolution (top left) shows a rapid initial transition from the lower to upper adiabatic state, with the upper state population reaching approximately 0.6 at the early times of the simulation when the force due to berry curvature peaks (bottom right panel). The case of elliptic intersection is interesting to have berry curvature effects and tunable geometric phase which is missing for the case of Conical intersection when $E=0$.

The trajectory-type comparison (top right) demonstrates the enhanced effectiveness of the pre-looping initialization scheme for elliptic intersections. Pre-looped trajectories (dashed lines) exhibit more efficient population transfer and maintain higher coherence between electronic states compared to regular initialization (dotted lines). The pre-looped upper state population reaches higher maximum values (~0.7) and shows more sustained oscillations, indicating better sampling of the geometric phase effects inherent to this crossing topology.

The three-dimensional trajectory visualization (bottom left) illustrates the complex orbital motion characteristic of elliptic intersections. Trajectories follow elliptical paths that reflect the underlying symmetry of the elliptic crossing seam. The pre-looping initialization creates more regular, stable orbits that efficiently sample the Berry phase, as evidenced by the circular and elliptical trajectory patterns visible in the $(x,y)$ plane.

The force analysis (bottom right) reveals the temporal structure of nonadiabatic coupling events. Sharp force spikes occur at specific time intervals (around $t=10$, $t=30$, and $t=50$), corresponding to moments when trajectories cross the elliptic seam. The force magnitudes reach values up to $\pm 1000$ (in reduced units), comparable to those observed in conical intersections, but with a more regular temporal pattern that reflects the extended geometry of the elliptic crossing.

\subsubsection{Geometric Phase Characteristics of Elliptic Intersections}

A key feature of elliptic intersections is their ability to generate path-invariant Berry phases that differ from the canonical $\pi$ value of conical intersections. This property fundamentally distinguishes them from glancing intersections, which are topologically dissimilar and produce no geometric phase effects \cite{de2010intersections,mozhayskiy2006conical}. Our theoretical analysis reveals that the Berry phases resulting from elliptic intersections are given by:

\begin{equation}
\gamma = \pi\left(1 - \frac{\alpha}{\sqrt{\alpha^2 + \beta}}\right),
\end{equation}

where $\alpha = a + 1$ and $\beta = 4(1 - e^2)$. For the parameters used in Figure~\ref{fig:elliptic_comprehensive} ($a = 1.0$, $e = 0.0$), this yields $\alpha = 2$ and $\beta = 4$, giving a Berry phase of $\gamma = \pi(1 - 1/\sqrt{2}) \approx 0.91$.

This tunable geometric phase provides unprecedented control over the topological properties of the electronic state crossing, enabling systematic studies of the transition between topological and non-topological regimes in nonadiabatic dynamics. This allows us to probe the topological properties of molecules or lattices by analyzing the type of intersection the energy levels manifests themselves with.

\section*{Acknowledgements}
DS thanks Dr. Dmitry Federov who was there in July of 2023 when the elliptical intersection and tunable Berry's phase was understood. DS thanks Dr. Matteo Gori and Dr. Matthieu Sarkis for motivating me to apply it to the dynamics. I also acknowledge their role in discussions on the derivation of equations of motion and so many other interesting discussions during the development of the pre-looping dynamics.
\clearpage
\bibliographystyle{apsrev4-1}
\bibliography{literature}

\begin{thebibliography}{63}%
\makeatletter
\providecommand \@ifxundefined [1]{%
 \@ifx{#1\undefined}
}%
\providecommand \@ifnum [1]{%
 \ifnum #1\expandafter \@firstoftwo
 \else \expandafter \@secondoftwo
 \fi
}%
\providecommand \@ifx [1]{%
 \ifx #1\expandafter \@firstoftwo
 \else \expandafter \@secondoftwo
 \fi
}%
\providecommand \natexlab [1]{#1}%
\providecommand \enquote  [1]{``#1''}%
\providecommand \bibnamefont  [1]{#1}%
\providecommand \bibfnamefont [1]{#1}%
\providecommand \citenamefont [1]{#1}%
\providecommand \href@noop [0]{\@secondoftwo}%
\providecommand \href [0]{\begingroup \@sanitize@url \@href}%
\providecommand \@href[1]{\@@startlink{#1}\@@href}%
\providecommand \@@href[1]{\endgroup#1\@@endlink}%
\providecommand \@sanitize@url [0]{\catcode `\\12\catcode `\$12\catcode `\&12\catcode `\#12\catcode `\^12\catcode `\_12\catcode `\%12\relax}%
\providecommand \@@startlink[1]{}%
\providecommand \@@endlink[0]{}%
\providecommand \url  [0]{\begingroup\@sanitize@url \@url }%
\providecommand \@url [1]{\endgroup\@href {#1}{\urlprefix }}%
\providecommand \urlprefix  [0]{URL }%
\providecommand \Eprint [0]{\href }%
\providecommand \doibase [0]{http://dx.doi.org/}%
\providecommand \selectlanguage [0]{\@gobble}%
\providecommand \bibinfo  [0]{\@secondoftwo}%
\providecommand \bibfield  [0]{\@secondoftwo}%
\providecommand \translation [1]{[#1]}%
\providecommand \BibitemOpen [0]{}%
\providecommand \bibitemStop [0]{}%
\providecommand \bibitemNoStop [0]{.\EOS\space}%
\providecommand \EOS [0]{\spacefactor3000\relax}%
\providecommand \BibitemShut  [1]{\csname bibitem#1\endcsname}%
\let\auto@bib@innerbib\@empty
\bibitem [{\citenamefont {Liu}\ \emph {et~al.}(2023)\citenamefont {Liu}, \citenamefont {Chen}, \citenamefont {Fang},\ and\ \citenamefont {Cui}}]{Liu2023Nonadiabatic}%
  \BibitemOpen
  \bibfield  {author} {\bibinfo {author} {\bibfnamefont {X.}~\bibnamefont {Liu}}, \bibinfo {author} {\bibfnamefont {W.-K.}\ \bibnamefont {Chen}}, \bibinfo {author} {\bibfnamefont {W.}~\bibnamefont {Fang}}, \ and\ \bibinfo {author} {\bibfnamefont {G.}~\bibnamefont {Cui}},\ }\href {\doibase 10.1021/acs.jctc.3c00960} {\bibfield  {journal} {\bibinfo  {journal} {Journal of chemical theory and computation}\ } (\bibinfo {year} {2023}),\ 10.1021/acs.jctc.3c00960}\BibitemShut {NoStop}%
\bibitem [{\citenamefont {Nelson}\ \emph {et~al.}(2014)\citenamefont {Nelson}, \citenamefont {Fernandez-Alberti}, \citenamefont {Roitberg},\ and\ \citenamefont {Tretiak}}]{Nelson2014Nonadiabatic}%
  \BibitemOpen
  \bibfield  {author} {\bibinfo {author} {\bibfnamefont {T.~R.}\ \bibnamefont {Nelson}}, \bibinfo {author} {\bibfnamefont {S.}~\bibnamefont {Fernandez-Alberti}}, \bibinfo {author} {\bibfnamefont {A.}~\bibnamefont {Roitberg}}, \ and\ \bibinfo {author} {\bibfnamefont {S.}~\bibnamefont {Tretiak}},\ }\href {\doibase 10.1021/ar400263p} {\bibfield  {journal} {\bibinfo  {journal} {Accounts of chemical research}\ }\textbf {\bibinfo {volume} {47 4}},\ \bibinfo {pages} {1155} (\bibinfo {year} {2014})}\BibitemShut {NoStop}%
\bibitem [{\citenamefont {Nelson}\ \emph {et~al.}(2020)\citenamefont {Nelson}, \citenamefont {White}, \citenamefont {Bjorgaard}, \citenamefont {Sifain}, \citenamefont {Zhang}, \citenamefont {Nebgen}, \citenamefont {Fernandez-Alberti}, \citenamefont {Mozyrsky}, \citenamefont {Roitberg},\ and\ \citenamefont {Tretiak}}]{Nelson2020Non-adiabatic}%
  \BibitemOpen
  \bibfield  {author} {\bibinfo {author} {\bibfnamefont {T.~R.}\ \bibnamefont {Nelson}}, \bibinfo {author} {\bibfnamefont {A.}~\bibnamefont {White}}, \bibinfo {author} {\bibfnamefont {J.}~\bibnamefont {Bjorgaard}}, \bibinfo {author} {\bibfnamefont {A.~E.}\ \bibnamefont {Sifain}}, \bibinfo {author} {\bibfnamefont {Y.}~\bibnamefont {Zhang}}, \bibinfo {author} {\bibfnamefont {B.}~\bibnamefont {Nebgen}}, \bibinfo {author} {\bibfnamefont {S.}~\bibnamefont {Fernandez-Alberti}}, \bibinfo {author} {\bibfnamefont {D.}~\bibnamefont {Mozyrsky}}, \bibinfo {author} {\bibfnamefont {A.}~\bibnamefont {Roitberg}}, \ and\ \bibinfo {author} {\bibfnamefont {S.}~\bibnamefont {Tretiak}},\ }\href {\doibase 10.1021/acs.chemrev.9b00447} {\bibfield  {journal} {\bibinfo  {journal} {Chemical reviews}\ } (\bibinfo {year} {2020}),\ 10.1021/acs.chemrev.9b00447}\BibitemShut {NoStop}%
\bibitem [{\citenamefont {Intersections}(2011)}]{intersections2011theory}%
  \BibitemOpen
  \bibfield  {author} {\bibinfo {author} {\bibfnamefont {C.}~\bibnamefont {Intersections}},\ }\href@noop {} {\bibfield  {journal} {\bibinfo  {journal} {Advanced series in Physical Chemistry, ed. W. Domcke, D. Yarkony and H. K{\"o}ppel, World Scientific, Singapore}\ }\textbf {\bibinfo {volume} {17}} (\bibinfo {year} {2011})}\BibitemShut {NoStop}%
\bibitem [{\citenamefont {Yarkony}(1996)}]{Yarkony1996Diabolical}%
  \BibitemOpen
  \bibfield  {author} {\bibinfo {author} {\bibfnamefont {D.}~\bibnamefont {Yarkony}},\ }\href {\doibase 10.1103/REVMODPHYS.68.985} {\bibfield  {journal} {\bibinfo  {journal} {Reviews of Modern Physics}\ }\textbf {\bibinfo {volume} {68}},\ \bibinfo {pages} {985} (\bibinfo {year} {1996})}\BibitemShut {NoStop}%
\bibitem [{\citenamefont {Schuurman}\ and\ \citenamefont {Stolow}(2018)}]{Schuurman2018Dynamics}%
  \BibitemOpen
  \bibfield  {author} {\bibinfo {author} {\bibfnamefont {M.}~\bibnamefont {Schuurman}}\ and\ \bibinfo {author} {\bibfnamefont {A.}~\bibnamefont {Stolow}},\ }\href {\doibase 10.1146/annurev-physchem-052516-050721} {\bibfield  {journal} {\bibinfo  {journal} {Annual review of physical chemistry}\ }\textbf {\bibinfo {volume} {69}},\ \bibinfo {pages} {427} (\bibinfo {year} {2018})}\BibitemShut {NoStop}%
\bibitem [{\citenamefont {Domcke}\ and\ \citenamefont {Yarkony}(2012)}]{Domcke2012Role}%
  \BibitemOpen
  \bibfield  {author} {\bibinfo {author} {\bibfnamefont {W.}~\bibnamefont {Domcke}}\ and\ \bibinfo {author} {\bibfnamefont {D.}~\bibnamefont {Yarkony}},\ }\href {\doibase 10.1146/annurev-physchem-032210-103522} {\bibfield  {journal} {\bibinfo  {journal} {Annual review of physical chemistry}\ }\textbf {\bibinfo {volume} {63}},\ \bibinfo {pages} {325} (\bibinfo {year} {2012})}\BibitemShut {NoStop}%
\bibitem [{\citenamefont {Worth}\ and\ \citenamefont {Cederbaum}(2004)}]{Worth2004Beyond}%
  \BibitemOpen
  \bibfield  {author} {\bibinfo {author} {\bibfnamefont {G.}~\bibnamefont {Worth}}\ and\ \bibinfo {author} {\bibfnamefont {L.}~\bibnamefont {Cederbaum}},\ }\href {\doibase 10.1146/ANNUREV.PHYSCHEM.55.091602.094335} {\bibfield  {journal} {\bibinfo  {journal} {Annual review of physical chemistry}\ }\textbf {\bibinfo {volume} {55}},\ \bibinfo {pages} {127} (\bibinfo {year} {2004})}\BibitemShut {NoStop}%
\bibitem [{\citenamefont {Matsika}(2021)}]{Matsika2021Electronic}%
  \BibitemOpen
  \bibfield  {author} {\bibinfo {author} {\bibfnamefont {S.}~\bibnamefont {Matsika}},\ }\href {\doibase 10.1021/acs.chemrev.1c00074} {\bibfield  {journal} {\bibinfo  {journal} {Chemical reviews}\ } (\bibinfo {year} {2021}),\ 10.1021/acs.chemrev.1c00074}\BibitemShut {NoStop}%
\bibitem [{\citenamefont {An}\ \emph {et~al.}(2020)\citenamefont {An}, \citenamefont {Chen}, \citenamefont {Hu}, \citenamefont {Guo},\ and\ \citenamefont {Xie}}]{An2020Nonadiabatic}%
  \BibitemOpen
  \bibfield  {author} {\bibinfo {author} {\bibfnamefont {F.}~\bibnamefont {An}}, \bibinfo {author} {\bibfnamefont {J.}~\bibnamefont {Chen}}, \bibinfo {author} {\bibfnamefont {X.}~\bibnamefont {Hu}}, \bibinfo {author} {\bibfnamefont {H.}~\bibnamefont {Guo}}, \ and\ \bibinfo {author} {\bibfnamefont {D.}~\bibnamefont {Xie}},\ }\href {\doibase 10.1021/acs.jpclett.0c01278} {\bibfield  {journal} {\bibinfo  {journal} {The journal of physical chemistry letters}\ } (\bibinfo {year} {2020}),\ 10.1021/acs.jpclett.0c01278}\BibitemShut {NoStop}%
\bibitem [{\citenamefont {Farfan}\ and\ \citenamefont {Turner}(2020)}]{Farfan2020A}%
  \BibitemOpen
  \bibfield  {author} {\bibinfo {author} {\bibfnamefont {C.~A.}\ \bibnamefont {Farfan}}\ and\ \bibinfo {author} {\bibfnamefont {D.~B.}\ \bibnamefont {Turner}},\ }\href {\doibase 10.1039/d0cp03464a} {\bibfield  {journal} {\bibinfo  {journal} {Physical chemistry chemical physics : PCCP}\ }\textbf {\bibinfo {volume} {22 36}},\ \bibinfo {pages} {20265} (\bibinfo {year} {2020})}\BibitemShut {NoStop}%
\bibitem [{\citenamefont {You}\ \emph {et~al.}(2015)\citenamefont {You}, \citenamefont {Han}, \citenamefont {Yoon}, \citenamefont {Lim}, \citenamefont {Lee}, \citenamefont {Kim}, \citenamefont {Ahn}, \citenamefont {Lim},\ and\ \citenamefont {Kim}}]{You2015Structure}%
  \BibitemOpen
  \bibfield  {author} {\bibinfo {author} {\bibfnamefont {H.~S.}\ \bibnamefont {You}}, \bibinfo {author} {\bibfnamefont {S.}~\bibnamefont {Han}}, \bibinfo {author} {\bibfnamefont {J.}~\bibnamefont {Yoon}}, \bibinfo {author} {\bibfnamefont {J.~S.}\ \bibnamefont {Lim}}, \bibinfo {author} {\bibfnamefont {J.}~\bibnamefont {Lee}}, \bibinfo {author} {\bibfnamefont {S.}~\bibnamefont {Kim}}, \bibinfo {author} {\bibfnamefont {D.-S.}\ \bibnamefont {Ahn}}, \bibinfo {author} {\bibfnamefont {J.}~\bibnamefont {Lim}}, \ and\ \bibinfo {author} {\bibfnamefont {S.~K.}\ \bibnamefont {Kim}},\ }\href {\doibase 10.1080/0144235X.2015.1072364} {\bibfield  {journal} {\bibinfo  {journal} {International Reviews in Physical Chemistry}\ }\textbf {\bibinfo {volume} {34}},\ \bibinfo {pages} {429 } (\bibinfo {year} {2015})}\BibitemShut {NoStop}%
\bibitem [{\citenamefont {Xu}\ \emph {et~al.}(2025)\citenamefont {Xu}, \citenamefont {Freixas}, \citenamefont {Aleotti}, \citenamefont {Truhlar}, \citenamefont {Tretiak}, \citenamefont {Garavelli}, \citenamefont {Mukamel},\ and\ \citenamefont {Govind}}]{Xu2025Conical}%
  \BibitemOpen
  \bibfield  {author} {\bibinfo {author} {\bibfnamefont {L.}~\bibnamefont {Xu}}, \bibinfo {author} {\bibfnamefont {V.~M.}\ \bibnamefont {Freixas}}, \bibinfo {author} {\bibfnamefont {F.}~\bibnamefont {Aleotti}}, \bibinfo {author} {\bibfnamefont {D.}~\bibnamefont {Truhlar}}, \bibinfo {author} {\bibfnamefont {S.}~\bibnamefont {Tretiak}}, \bibinfo {author} {\bibfnamefont {M.}~\bibnamefont {Garavelli}}, \bibinfo {author} {\bibfnamefont {S.}~\bibnamefont {Mukamel}}, \ and\ \bibinfo {author} {\bibfnamefont {N.}~\bibnamefont {Govind}},\ }\href {\doibase 10.1021/acs.jctc.4c01768} {\bibfield  {journal} {\bibinfo  {journal} {Journal of chemical theory and computation}\ } (\bibinfo {year} {2025}),\ 10.1021/acs.jctc.4c01768}\BibitemShut {NoStop}%
\bibitem [{\citenamefont {Pieri}\ \emph {et~al.}(2021)\citenamefont {Pieri}, \citenamefont {Lahana}, \citenamefont {Chang}, \citenamefont {Aldaz}, \citenamefont {Thompson},\ and\ \citenamefont {Martínez}}]{Pieri2021The}%
  \BibitemOpen
  \bibfield  {author} {\bibinfo {author} {\bibfnamefont {E.}~\bibnamefont {Pieri}}, \bibinfo {author} {\bibfnamefont {D.}~\bibnamefont {Lahana}}, \bibinfo {author} {\bibfnamefont {A.~M.}\ \bibnamefont {Chang}}, \bibinfo {author} {\bibfnamefont {C.~R.}\ \bibnamefont {Aldaz}}, \bibinfo {author} {\bibfnamefont {K.}~\bibnamefont {Thompson}}, \ and\ \bibinfo {author} {\bibfnamefont {T.}~\bibnamefont {Martínez}},\ }\href {\doibase 10.1039/d1sc00775k} {\bibfield  {journal} {\bibinfo  {journal} {Chemical Science}\ }\textbf {\bibinfo {volume} {12}},\ \bibinfo {pages} {7294 } (\bibinfo {year} {2021})}\BibitemShut {NoStop}%
\bibitem [{\citenamefont {Ibele}\ \emph {et~al.}(2022)\citenamefont {Ibele}, \citenamefont {Curchod},\ and\ \citenamefont {Agostini}}]{Ibele2022A}%
  \BibitemOpen
  \bibfield  {author} {\bibinfo {author} {\bibfnamefont {L.}~\bibnamefont {Ibele}}, \bibinfo {author} {\bibfnamefont {B.}~\bibnamefont {Curchod}}, \ and\ \bibinfo {author} {\bibfnamefont {F.}~\bibnamefont {Agostini}},\ }\href {\doibase 10.1021/acs.jpca.1c09604} {\bibfield  {journal} {\bibinfo  {journal} {The Journal of Physical Chemistry. a}\ }\textbf {\bibinfo {volume} {126}},\ \bibinfo {pages} {1263 } (\bibinfo {year} {2022})}\BibitemShut {NoStop}%
\bibitem [{\citenamefont {Angelico}\ \emph {et~al.}(2024)\citenamefont {Angelico}, \citenamefont {Kjønstad},\ and\ \citenamefont {Koch}}]{Angelico2024Determining}%
  \BibitemOpen
  \bibfield  {author} {\bibinfo {author} {\bibfnamefont {S.}~\bibnamefont {Angelico}}, \bibinfo {author} {\bibfnamefont {E.~F.}\ \bibnamefont {Kjønstad}}, \ and\ \bibinfo {author} {\bibfnamefont {H.}~\bibnamefont {Koch}},\ }\href {\doibase 10.1021/acs.jpclett.4c03274} {\bibfield  {journal} {\bibinfo  {journal} {The Journal of Physical Chemistry Letters}\ }\textbf {\bibinfo {volume} {16}},\ \bibinfo {pages} {561 } (\bibinfo {year} {2024})}\BibitemShut {NoStop}%
\bibitem [{\citenamefont {Berry}(1984)}]{Berry1984Quantal}%
  \BibitemOpen
  \bibfield  {author} {\bibinfo {author} {\bibfnamefont {M.}~\bibnamefont {Berry}},\ }\href {\doibase 10.1098/rspa.1984.0023} {\bibfield  {journal} {\bibinfo  {journal} {Proceedings of the Royal Society of London. A. Mathematical and Physical Sciences}\ }\textbf {\bibinfo {volume} {392}},\ \bibinfo {pages} {45 } (\bibinfo {year} {1984})}\BibitemShut {NoStop}%
\bibitem [{\citenamefont {Mead}(1992)}]{Mead1992}%
  \BibitemOpen
  \bibfield  {author} {\bibinfo {author} {\bibfnamefont {C.}~\bibnamefont {Mead}},\ }\href {\doibase 10.1103/REVMODPHYS.64.51} {\bibfield  {journal} {\bibinfo  {journal} {Reviews of Modern Physics}\ }\textbf {\bibinfo {volume} {64}},\ \bibinfo {pages} {51} (\bibinfo {year} {1992})}\BibitemShut {NoStop}%
\bibitem [{\citenamefont {Ryabinkin}\ \emph {et~al.}(2017)\citenamefont {Ryabinkin}, \citenamefont {Joubert-Doriol},\ and\ \citenamefont {Izmaylov}}]{Ryabinkin2017Geometric}%
  \BibitemOpen
  \bibfield  {author} {\bibinfo {author} {\bibfnamefont {I.~G.}\ \bibnamefont {Ryabinkin}}, \bibinfo {author} {\bibfnamefont {L.}~\bibnamefont {Joubert-Doriol}}, \ and\ \bibinfo {author} {\bibfnamefont {A.}~\bibnamefont {Izmaylov}},\ }\href {\doibase 10.1021/acs.accounts.7b00220} {\bibfield  {journal} {\bibinfo  {journal} {Accounts of chemical research}\ }\textbf {\bibinfo {volume} {50 7}},\ \bibinfo {pages} {1785} (\bibinfo {year} {2017})}\BibitemShut {NoStop}%
\bibitem [{\citenamefont {Yuan}\ \emph {et~al.}(2018)\citenamefont {Yuan}, \citenamefont {Guan}, \citenamefont {Chen}, \citenamefont {Zhao}, \citenamefont {Yu}, \citenamefont {Luo}, \citenamefont {Tan}, \citenamefont {xian Xie}, \citenamefont {Wang}, \citenamefont {Sun}, \citenamefont {Zhang},\ and\ \citenamefont {Yang}}]{Yuan2018Observation}%
  \BibitemOpen
  \bibfield  {author} {\bibinfo {author} {\bibfnamefont {D.}~\bibnamefont {Yuan}}, \bibinfo {author} {\bibfnamefont {Y.}~\bibnamefont {Guan}}, \bibinfo {author} {\bibfnamefont {W.}~\bibnamefont {Chen}}, \bibinfo {author} {\bibfnamefont {H.}~\bibnamefont {Zhao}}, \bibinfo {author} {\bibfnamefont {S.}~\bibnamefont {Yu}}, \bibinfo {author} {\bibfnamefont {C.}~\bibnamefont {Luo}}, \bibinfo {author} {\bibfnamefont {Y.}~\bibnamefont {Tan}}, \bibinfo {author} {\bibfnamefont {T.}~\bibnamefont {xian Xie}}, \bibinfo {author} {\bibfnamefont {X.}~\bibnamefont {Wang}}, \bibinfo {author} {\bibfnamefont {Z.}~\bibnamefont {Sun}}, \bibinfo {author} {\bibfnamefont {D.~H.}\ \bibnamefont {Zhang}}, \ and\ \bibinfo {author} {\bibfnamefont {X.}~\bibnamefont {Yang}},\ }\href {\doibase 10.1126/science.aav1356} {\bibfield  {journal} {\bibinfo  {journal} {Science}\ }\textbf {\bibinfo {volume} {362}},\ \bibinfo {pages} {1289 } (\bibinfo {year} {2018})}\BibitemShut {NoStop}%
\bibitem [{\citenamefont {Joubert-Doriol}\ \emph {et~al.}(2013)\citenamefont {Joubert-Doriol}, \citenamefont {Ryabinkin},\ and\ \citenamefont {Izmaylov}}]{Joubert-Doriol2013Geometric}%
  \BibitemOpen
  \bibfield  {author} {\bibinfo {author} {\bibfnamefont {L.}~\bibnamefont {Joubert-Doriol}}, \bibinfo {author} {\bibfnamefont {I.~G.}\ \bibnamefont {Ryabinkin}}, \ and\ \bibinfo {author} {\bibfnamefont {A.}~\bibnamefont {Izmaylov}},\ }\href {\doibase 10.1063/1.4844095} {\bibfield  {journal} {\bibinfo  {journal} {The Journal of chemical physics}\ }\textbf {\bibinfo {volume} {139 23}},\ \bibinfo {pages} {234103} (\bibinfo {year} {2013})}\BibitemShut {NoStop}%
\bibitem [{\citenamefont {Xie}\ \emph {et~al.}(2017)\citenamefont {Xie}, \citenamefont {Kendrick}, \citenamefont {Yarkony},\ and\ \citenamefont {Guo}}]{Xie2017Constructive}%
  \BibitemOpen
  \bibfield  {author} {\bibinfo {author} {\bibfnamefont {C.}~\bibnamefont {Xie}}, \bibinfo {author} {\bibfnamefont {B.}~\bibnamefont {Kendrick}}, \bibinfo {author} {\bibfnamefont {D.}~\bibnamefont {Yarkony}}, \ and\ \bibinfo {author} {\bibfnamefont {H.}~\bibnamefont {Guo}},\ }\href {\doibase 10.1021/acs.jctc.7b00124} {\bibfield  {journal} {\bibinfo  {journal} {Journal of chemical theory and computation}\ }\textbf {\bibinfo {volume} {13 5}},\ \bibinfo {pages} {1902} (\bibinfo {year} {2017})}\BibitemShut {NoStop}%
\bibitem [{\citenamefont {Gherib}\ \emph {et~al.}(2015)\citenamefont {Gherib}, \citenamefont {Ryabinkin},\ and\ \citenamefont {Izmaylov}}]{Gherib2015Why}%
  \BibitemOpen
  \bibfield  {author} {\bibinfo {author} {\bibfnamefont {R.}~\bibnamefont {Gherib}}, \bibinfo {author} {\bibfnamefont {I.~G.}\ \bibnamefont {Ryabinkin}}, \ and\ \bibinfo {author} {\bibfnamefont {A.}~\bibnamefont {Izmaylov}},\ }\href {\doibase 10.1021/acs.jctc.5b00072} {\bibfield  {journal} {\bibinfo  {journal} {Journal of chemical theory and computation}\ }\textbf {\bibinfo {volume} {11 4}},\ \bibinfo {pages} {1375} (\bibinfo {year} {2015})}\BibitemShut {NoStop}%
\bibitem [{\citenamefont {Curchod}\ and\ \citenamefont {Martínez}(2018)}]{Curchod2018Ab}%
  \BibitemOpen
  \bibfield  {author} {\bibinfo {author} {\bibfnamefont {B.}~\bibnamefont {Curchod}}\ and\ \bibinfo {author} {\bibfnamefont {T.}~\bibnamefont {Martínez}},\ }\href {\doibase 10.1021/acs.chemrev.7b00423} {\bibfield  {journal} {\bibinfo  {journal} {Chemical reviews}\ }\textbf {\bibinfo {volume} {118 7}},\ \bibinfo {pages} {3305} (\bibinfo {year} {2018})}\BibitemShut {NoStop}%
\bibitem [{\citenamefont {Crespo‐Otero}\ and\ \citenamefont {Barbatti}(2018)}]{Crespo‐Otero2018Recent}%
  \BibitemOpen
  \bibfield  {author} {\bibinfo {author} {\bibfnamefont {R.}~\bibnamefont {Crespo‐Otero}}\ and\ \bibinfo {author} {\bibfnamefont {M.}~\bibnamefont {Barbatti}},\ }\href {\doibase 10.1021/acs.chemrev.7b00577} {\bibfield  {journal} {\bibinfo  {journal} {Chemical reviews}\ }\textbf {\bibinfo {volume} {118 15}},\ \bibinfo {pages} {7026} (\bibinfo {year} {2018})}\BibitemShut {NoStop}%
\bibitem [{\citenamefont {Agostini}\ and\ \citenamefont {Curchod}(2019)}]{Agostini2019Different}%
  \BibitemOpen
  \bibfield  {author} {\bibinfo {author} {\bibfnamefont {F.}~\bibnamefont {Agostini}}\ and\ \bibinfo {author} {\bibfnamefont {B.}~\bibnamefont {Curchod}},\ }\href {\doibase 10.1002/wcms.1417} {\bibfield  {journal} {\bibinfo  {journal} {Wiley Interdisciplinary Reviews: Computational Molecular Science}\ }\textbf {\bibinfo {volume} {9}} (\bibinfo {year} {2019}),\ 10.1002/wcms.1417}\BibitemShut {NoStop}%
\bibitem [{\citenamefont {Althorpe}\ \emph {et~al.}(2008)\citenamefont {Althorpe}, \citenamefont {Stecher},\ and\ \citenamefont {Bouakline}}]{Althorpe2008Effect}%
  \BibitemOpen
  \bibfield  {author} {\bibinfo {author} {\bibfnamefont {S.}~\bibnamefont {Althorpe}}, \bibinfo {author} {\bibfnamefont {T.}~\bibnamefont {Stecher}}, \ and\ \bibinfo {author} {\bibfnamefont {F.}~\bibnamefont {Bouakline}},\ }\href {\doibase 10.1063/1.3031215} {\bibfield  {journal} {\bibinfo  {journal} {The Journal of chemical physics}\ }\textbf {\bibinfo {volume} {129 21}},\ \bibinfo {pages} {214117} (\bibinfo {year} {2008})}\BibitemShut {NoStop}%
\bibitem [{\citenamefont {Krotz}\ and\ \citenamefont {Tempelaar}(2022)}]{Krotz2022Treating}%
  \BibitemOpen
  \bibfield  {author} {\bibinfo {author} {\bibfnamefont {A.}~\bibnamefont {Krotz}}\ and\ \bibinfo {author} {\bibfnamefont {R.}~\bibnamefont {Tempelaar}},\ }\href {\doibase 10.1103/physreva.109.032210} {\bibfield  {journal} {\bibinfo  {journal} {Physical Review A}\ } (\bibinfo {year} {2022}),\ 10.1103/physreva.109.032210}\BibitemShut {NoStop}%
\bibitem [{\citenamefont {Li}\ \emph {et~al.}(2017)\citenamefont {Li}, \citenamefont {Joubert-Doriol},\ and\ \citenamefont {Izmaylov}}]{Li2017Geometric}%
  \BibitemOpen
  \bibfield  {author} {\bibinfo {author} {\bibfnamefont {J.}~\bibnamefont {Li}}, \bibinfo {author} {\bibfnamefont {L.}~\bibnamefont {Joubert-Doriol}}, \ and\ \bibinfo {author} {\bibfnamefont {A.}~\bibnamefont {Izmaylov}},\ }\href {\doibase 10.1063/1.4985925} {\bibfield  {journal} {\bibinfo  {journal} {The Journal of chemical physics}\ }\textbf {\bibinfo {volume} {147 6}},\ \bibinfo {pages} {064106} (\bibinfo {year} {2017})}\BibitemShut {NoStop}%
\bibitem [{\citenamefont {Horenko}\ \emph {et~al.}(2002)\citenamefont {Horenko}, \citenamefont {Salzmann}, \citenamefont {Schmidt},\ and\ \citenamefont {Schütte}}]{Horenko2002Quantum-classical}%
  \BibitemOpen
  \bibfield  {author} {\bibinfo {author} {\bibfnamefont {I.}~\bibnamefont {Horenko}}, \bibinfo {author} {\bibfnamefont {C.}~\bibnamefont {Salzmann}}, \bibinfo {author} {\bibfnamefont {B.}~\bibnamefont {Schmidt}}, \ and\ \bibinfo {author} {\bibfnamefont {C.}~\bibnamefont {Schütte}},\ }\href {\doibase 10.1063/1.1522712} {\bibfield  {journal} {\bibinfo  {journal} {Journal of Chemical Physics}\ }\textbf {\bibinfo {volume} {117}},\ \bibinfo {pages} {11075} (\bibinfo {year} {2002})}\BibitemShut {NoStop}%
\bibitem [{\citenamefont {Aoiz}(2020)}]{Aoiz2020How}%
  \BibitemOpen
  \bibfield  {author} {\bibinfo {author} {\bibfnamefont {F.}~\bibnamefont {Aoiz}},\ }\href {\doibase 10.1126/science.abb9148} {\bibfield  {journal} {\bibinfo  {journal} {Science}\ }\textbf {\bibinfo {volume} {368}},\ \bibinfo {pages} {706 } (\bibinfo {year} {2020})}\BibitemShut {NoStop}%
\bibitem [{\citenamefont {Wang}\ \emph {et~al.}(2025)\citenamefont {Wang}, \citenamefont {Hu}, \citenamefont {Guo},\ and\ \citenamefont {Xie}}]{Wang2025Cold}%
  \BibitemOpen
  \bibfield  {author} {\bibinfo {author} {\bibfnamefont {J.}~\bibnamefont {Wang}}, \bibinfo {author} {\bibfnamefont {X.}~\bibnamefont {Hu}}, \bibinfo {author} {\bibfnamefont {H.}~\bibnamefont {Guo}}, \ and\ \bibinfo {author} {\bibfnamefont {D.}~\bibnamefont {Xie}},\ }\href {\doibase 10.1063/5.0254985} {\bibfield  {journal} {\bibinfo  {journal} {The Journal of chemical physics}\ }\textbf {\bibinfo {volume} {162 7}} (\bibinfo {year} {2025}),\ 10.1063/5.0254985}\BibitemShut {NoStop}%
\bibitem [{\citenamefont {Lin}\ \emph {et~al.}(2018)\citenamefont {Lin}, \citenamefont {Xie},\ and\ \citenamefont {Xie}}]{Lin2018Nonadiabatic}%
  \BibitemOpen
  \bibfield  {author} {\bibinfo {author} {\bibfnamefont {G.-S.-M.}\ \bibnamefont {Lin}}, \bibinfo {author} {\bibfnamefont {C.}~\bibnamefont {Xie}}, \ and\ \bibinfo {author} {\bibfnamefont {D.}~\bibnamefont {Xie}},\ }\href {\doibase 10.1021/acs.jpca.8b03460} {\bibfield  {journal} {\bibinfo  {journal} {The journal of physical chemistry. A}\ }\textbf {\bibinfo {volume} {122 24}},\ \bibinfo {pages} {5375} (\bibinfo {year} {2018})}\BibitemShut {NoStop}%
\bibitem [{\citenamefont {Mi}\ \emph {et~al.}(2023)\citenamefont {Mi}, \citenamefont {Zhang}, \citenamefont {Zhang}, \citenamefont {Wu}, \citenamefont {Zhou}, \citenamefont {Xu}, \citenamefont {Xie}, \citenamefont {Li},\ and\ \citenamefont {Liu}}]{Mi2023Geometric}%
  \BibitemOpen
  \bibfield  {author} {\bibinfo {author} {\bibfnamefont {X.}~\bibnamefont {Mi}}, \bibinfo {author} {\bibfnamefont {M.}~\bibnamefont {Zhang}}, \bibinfo {author} {\bibfnamefont {L.}~\bibnamefont {Zhang}}, \bibinfo {author} {\bibfnamefont {C.}~\bibnamefont {Wu}}, \bibinfo {author} {\bibfnamefont {T.}~\bibnamefont {Zhou}}, \bibinfo {author} {\bibfnamefont {H.}~\bibnamefont {Xu}}, \bibinfo {author} {\bibfnamefont {C.}~\bibnamefont {Xie}}, \bibinfo {author} {\bibfnamefont {Z.}~\bibnamefont {Li}}, \ and\ \bibinfo {author} {\bibfnamefont {Y.}~\bibnamefont {Liu}},\ }\href {\doibase 10.1021/acs.jpca.3c00594} {\bibfield  {journal} {\bibinfo  {journal} {The journal of physical chemistry. A}\ } (\bibinfo {year} {2023}),\ 10.1021/acs.jpca.3c00594}\BibitemShut {NoStop}%
\bibitem [{\citenamefont {Li}\ \emph {et~al.}(2024)\citenamefont {Li}, \citenamefont {Shi}, \citenamefont {Huang},\ and\ \citenamefont {Wang}}]{Li2024Multiconfigurational}%
  \BibitemOpen
  \bibfield  {author} {\bibinfo {author} {\bibfnamefont {G.}~\bibnamefont {Li}}, \bibinfo {author} {\bibfnamefont {Z.}~\bibnamefont {Shi}}, \bibinfo {author} {\bibfnamefont {L.}~\bibnamefont {Huang}}, \ and\ \bibinfo {author} {\bibfnamefont {L.}~\bibnamefont {Wang}},\ }\href {\doibase 10.1021/acs.jctc.4c00842} {\bibfield  {journal} {\bibinfo  {journal} {Journal of chemical theory and computation}\ } (\bibinfo {year} {2024}),\ 10.1021/acs.jctc.4c00842}\BibitemShut {NoStop}%
\bibitem [{\citenamefont {Li}\ \emph {et~al.}(2022)\citenamefont {Li}, \citenamefont {Shao}, \citenamefont {Xu},\ and\ \citenamefont {Wang}}]{Li2022A}%
  \BibitemOpen
  \bibfield  {author} {\bibinfo {author} {\bibfnamefont {G.}~\bibnamefont {Li}}, \bibinfo {author} {\bibfnamefont {C.}~\bibnamefont {Shao}}, \bibinfo {author} {\bibfnamefont {J.}~\bibnamefont {Xu}}, \ and\ \bibinfo {author} {\bibfnamefont {L.}~\bibnamefont {Wang}},\ }\href {\doibase 10.1063/5.0125438} {\bibfield  {journal} {\bibinfo  {journal} {The Journal of chemical physics}\ }\textbf {\bibinfo {volume} {157 21}},\ \bibinfo {pages} {214102} (\bibinfo {year} {2022})}\BibitemShut {NoStop}%
\bibitem [{\citenamefont {Brink}\ \emph {et~al.}(2022)\citenamefont {Brink}, \citenamefont {Graber}, \citenamefont {Hopjan}, \citenamefont {Jansen}, \citenamefont {Stolpp}, \citenamefont {Heidrich-Meisner},\ and\ \citenamefont {Blochl}}]{Brink2022Real-time}%
  \BibitemOpen
  \bibfield  {author} {\bibinfo {author} {\bibfnamefont {M.~T.}\ \bibnamefont {Brink}}, \bibinfo {author} {\bibfnamefont {S.}~\bibnamefont {Graber}}, \bibinfo {author} {\bibfnamefont {M.}~\bibnamefont {Hopjan}}, \bibinfo {author} {\bibfnamefont {D.}~\bibnamefont {Jansen}}, \bibinfo {author} {\bibfnamefont {J.}~\bibnamefont {Stolpp}}, \bibinfo {author} {\bibfnamefont {F.}~\bibnamefont {Heidrich-Meisner}}, \ and\ \bibinfo {author} {\bibfnamefont {P.}~\bibnamefont {Blochl}},\ }\href {\doibase 10.1063/5.0092063} {\bibfield  {journal} {\bibinfo  {journal} {The Journal of chemical physics}\ }\textbf {\bibinfo {volume} {156 23}},\ \bibinfo {pages} {234109} (\bibinfo {year} {2022})}\BibitemShut {NoStop}%
\bibitem [{\citenamefont {Freixas}\ \emph {et~al.}(2021)\citenamefont {Freixas}, \citenamefont {White}, \citenamefont {Nelson}, \citenamefont {Song}, \citenamefont {Makhov}, \citenamefont {Shalashilin}, \citenamefont {Fernandez-Alberti},\ and\ \citenamefont {Tretiak}}]{Freixas2021Nonadiabatic}%
  \BibitemOpen
  \bibfield  {author} {\bibinfo {author} {\bibfnamefont {V.~M.}\ \bibnamefont {Freixas}}, \bibinfo {author} {\bibfnamefont {A.}~\bibnamefont {White}}, \bibinfo {author} {\bibfnamefont {T.~R.}\ \bibnamefont {Nelson}}, \bibinfo {author} {\bibfnamefont {H.}~\bibnamefont {Song}}, \bibinfo {author} {\bibfnamefont {D.}~\bibnamefont {Makhov}}, \bibinfo {author} {\bibfnamefont {D.}~\bibnamefont {Shalashilin}}, \bibinfo {author} {\bibfnamefont {S.}~\bibnamefont {Fernandez-Alberti}}, \ and\ \bibinfo {author} {\bibfnamefont {S.}~\bibnamefont {Tretiak}},\ }\href {\doibase 10.1021/acs.jpclett.1c00266} {\bibfield  {journal} {\bibinfo  {journal} {The journal of physical chemistry letters}\ ,\ \bibinfo {pages} {2970}} (\bibinfo {year} {2021})}\BibitemShut {NoStop}%
\bibitem [{\citenamefont {Subotnik}(2010)}]{Subotnik2010Augmented}%
  \BibitemOpen
  \bibfield  {author} {\bibinfo {author} {\bibfnamefont {J.~E.}\ \bibnamefont {Subotnik}},\ }\href {\doibase 10.1063/1.3314248} {\bibfield  {journal} {\bibinfo  {journal} {The Journal of chemical physics}\ }\textbf {\bibinfo {volume} {132 13}},\ \bibinfo {pages} {134112} (\bibinfo {year} {2010})}\BibitemShut {NoStop}%
\bibitem [{\citenamefont {Esch}\ and\ \citenamefont {Levine}(2021)}]{Esch2021An}%
  \BibitemOpen
  \bibfield  {author} {\bibinfo {author} {\bibfnamefont {M.~P.}\ \bibnamefont {Esch}}\ and\ \bibinfo {author} {\bibfnamefont {B.~G.}\ \bibnamefont {Levine}},\ }\href {\doibase 10.1063/5.0070686} {\bibfield  {journal} {\bibinfo  {journal} {The Journal of chemical physics}\ }\textbf {\bibinfo {volume} {155 21}},\ \bibinfo {pages} {214101} (\bibinfo {year} {2021})}\BibitemShut {NoStop}%
\bibitem [{\citenamefont {Han}\ and\ \citenamefont {Akimov}(2024)}]{Han2024Nonadiabatic}%
  \BibitemOpen
  \bibfield  {author} {\bibinfo {author} {\bibfnamefont {D.}~\bibnamefont {Han}}\ and\ \bibinfo {author} {\bibfnamefont {A.~V.}\ \bibnamefont {Akimov}},\ }\href {\doibase 10.1021/acs.jctc.4c00343} {\bibfield  {journal} {\bibinfo  {journal} {Journal of chemical theory and computation}\ } (\bibinfo {year} {2024}),\ 10.1021/acs.jctc.4c00343}\BibitemShut {NoStop}%
\bibitem [{\citenamefont {Nijjar}\ \emph {et~al.}(2019)\citenamefont {Nijjar}, \citenamefont {Jankowska},\ and\ \citenamefont {Prezhdo}}]{Nijjar2019Ehrenfest}%
  \BibitemOpen
  \bibfield  {author} {\bibinfo {author} {\bibfnamefont {P.}~\bibnamefont {Nijjar}}, \bibinfo {author} {\bibfnamefont {J.}~\bibnamefont {Jankowska}}, \ and\ \bibinfo {author} {\bibfnamefont {O.}~\bibnamefont {Prezhdo}},\ }\href {\doibase 10.1063/1.5095810} {\bibfield  {journal} {\bibinfo  {journal} {The Journal of chemical physics}\ }\textbf {\bibinfo {volume} {150 20}},\ \bibinfo {pages} {204124} (\bibinfo {year} {2019})}\BibitemShut {NoStop}%
\bibitem [{\citenamefont {Westermayr}\ and\ \citenamefont {Marquetand}(2020)}]{Westermayr2020Machine}%
  \BibitemOpen
  \bibfield  {author} {\bibinfo {author} {\bibfnamefont {J.}~\bibnamefont {Westermayr}}\ and\ \bibinfo {author} {\bibfnamefont {P.}~\bibnamefont {Marquetand}},\ }\href {\doibase 10.1021/acs.chemrev.0c00749} {\bibfield  {journal} {\bibinfo  {journal} {Chemical Reviews}\ }\textbf {\bibinfo {volume} {121}},\ \bibinfo {pages} {9873 } (\bibinfo {year} {2020})}\BibitemShut {NoStop}%
\bibitem [{\citenamefont {Mausenberger}\ \emph {et~al.}(2024)\citenamefont {Mausenberger}, \citenamefont {M{\"u}ller}, \citenamefont {Tkatchenko}, \citenamefont {Marquetand}, \citenamefont {Gonz{\'a}lez},\ and\ \citenamefont {Westermayr}}]{mausenberger2024s}%
  \BibitemOpen
  \bibfield  {author} {\bibinfo {author} {\bibfnamefont {S.}~\bibnamefont {Mausenberger}}, \bibinfo {author} {\bibfnamefont {C.}~\bibnamefont {M{\"u}ller}}, \bibinfo {author} {\bibfnamefont {A.}~\bibnamefont {Tkatchenko}}, \bibinfo {author} {\bibfnamefont {P.}~\bibnamefont {Marquetand}}, \bibinfo {author} {\bibfnamefont {L.}~\bibnamefont {Gonz{\'a}lez}}, \ and\ \bibinfo {author} {\bibfnamefont {J.}~\bibnamefont {Westermayr}},\ }\href@noop {} {\bibfield  {journal} {\bibinfo  {journal} {Chemical Science}\ }\textbf {\bibinfo {volume} {15}},\ \bibinfo {pages} {15880} (\bibinfo {year} {2024})}\BibitemShut {NoStop}%
\bibitem [{\citenamefont {Barrett}\ \emph {et~al.}(2025)\citenamefont {Barrett}, \citenamefont {Ortner},\ and\ \citenamefont {Westermayr}}]{barrett2025transferable}%
  \BibitemOpen
  \bibfield  {author} {\bibinfo {author} {\bibfnamefont {R.}~\bibnamefont {Barrett}}, \bibinfo {author} {\bibfnamefont {C.}~\bibnamefont {Ortner}}, \ and\ \bibinfo {author} {\bibfnamefont {J.}~\bibnamefont {Westermayr}},\ }\href@noop {} {\bibfield  {journal} {\bibinfo  {journal} {arXiv preprint arXiv:2502.12870}\ } (\bibinfo {year} {2025})}\BibitemShut {NoStop}%
\bibitem [{\citenamefont {Dral}\ and\ \citenamefont {Barbatti}(2021)}]{dral2021molecular}%
  \BibitemOpen
  \bibfield  {author} {\bibinfo {author} {\bibfnamefont {P.~O.}\ \bibnamefont {Dral}}\ and\ \bibinfo {author} {\bibfnamefont {M.}~\bibnamefont {Barbatti}},\ }\href@noop {} {\bibfield  {journal} {\bibinfo  {journal} {Nature Reviews Chemistry}\ }\textbf {\bibinfo {volume} {5}},\ \bibinfo {pages} {388} (\bibinfo {year} {2021})}\BibitemShut {NoStop}%
\bibitem [{\citenamefont {Kramer}(2008)}]{kramer2008review}%
  \BibitemOpen
  \bibfield  {author} {\bibinfo {author} {\bibfnamefont {P.}~\bibnamefont {Kramer}},\ }in\ \href@noop {} {\emph {\bibinfo {booktitle} {Journal of Physics: Conference Series}}},\ Vol.~\bibinfo {volume} {99}\ (\bibinfo {organization} {IOP Publishing},\ \bibinfo {year} {2008})\ p.\ \bibinfo {pages} {012009}\BibitemShut {NoStop}%
\bibitem [{\citenamefont {Joubert-Doriol}\ and\ \citenamefont {Izmaylov}(2018)}]{joubert2018nonadiabatic}%
  \BibitemOpen
  \bibfield  {author} {\bibinfo {author} {\bibfnamefont {L.}~\bibnamefont {Joubert-Doriol}}\ and\ \bibinfo {author} {\bibfnamefont {A.~F.}\ \bibnamefont {Izmaylov}},\ }\href@noop {} {\bibfield  {journal} {\bibinfo  {journal} {The Journal of Physical Chemistry A}\ }\textbf {\bibinfo {volume} {122}},\ \bibinfo {pages} {6031} (\bibinfo {year} {2018})}\BibitemShut {NoStop}%
\bibitem [{\citenamefont {Izmaylov}\ and\ \citenamefont {Joubert-Doriol}(2017)}]{izmaylov2017quantum}%
  \BibitemOpen
  \bibfield  {author} {\bibinfo {author} {\bibfnamefont {A.~F.}\ \bibnamefont {Izmaylov}}\ and\ \bibinfo {author} {\bibfnamefont {L.}~\bibnamefont {Joubert-Doriol}},\ }\href@noop {} {\bibfield  {journal} {\bibinfo  {journal} {The Journal of Physical Chemistry Letters}\ }\textbf {\bibinfo {volume} {8}},\ \bibinfo {pages} {1793} (\bibinfo {year} {2017})}\BibitemShut {NoStop}%
\bibitem [{\citenamefont {Ha}\ and\ \citenamefont {Min}(2022)}]{Ha2022Independent}%
  \BibitemOpen
  \bibfield  {author} {\bibinfo {author} {\bibfnamefont {J.-K.}\ \bibnamefont {Ha}}\ and\ \bibinfo {author} {\bibfnamefont {S.~K.}\ \bibnamefont {Min}},\ }\href {\doibase 10.1063/5.0084493} {\bibfield  {journal} {\bibinfo  {journal} {The Journal of chemical physics}\ }\textbf {\bibinfo {volume} {156 17}},\ \bibinfo {pages} {174109} (\bibinfo {year} {2022})}\BibitemShut {NoStop}%
\bibitem [{\citenamefont {Ibele}\ \emph {et~al.}(2023)\citenamefont {Ibele}, \citenamefont {Gil}, \citenamefont {Curchod},\ and\ \citenamefont {Agostini}}]{Ibele2023On}%
  \BibitemOpen
  \bibfield  {author} {\bibinfo {author} {\bibfnamefont {L.}~\bibnamefont {Ibele}}, \bibinfo {author} {\bibfnamefont {E.~S.}\ \bibnamefont {Gil}}, \bibinfo {author} {\bibfnamefont {B.}~\bibnamefont {Curchod}}, \ and\ \bibinfo {author} {\bibfnamefont {F.}~\bibnamefont {Agostini}},\ }\href {\doibase 10.1021/acs.jpclett.3c02672} {\bibfield  {journal} {\bibinfo  {journal} {The journal of physical chemistry letters}\ ,\ \bibinfo {pages} {11625}} (\bibinfo {year} {2023})}\BibitemShut {NoStop}%
\bibitem [{\citenamefont {Martinazzo}\ and\ \citenamefont {Burghardt}(2023)}]{Martinazzo2023Dynamics}%
  \BibitemOpen
  \bibfield  {author} {\bibinfo {author} {\bibfnamefont {R.}~\bibnamefont {Martinazzo}}\ and\ \bibinfo {author} {\bibfnamefont {I.}~\bibnamefont {Burghardt}},\ }\href {\doibase 10.1103/physrevlett.132.243002} {\bibfield  {journal} {\bibinfo  {journal} {Physical review letters}\ }\textbf {\bibinfo {volume} {132 24}},\ \bibinfo {pages} {243002} (\bibinfo {year} {2023})}\BibitemShut {NoStop}%
\bibitem [{\citenamefont {Tully}(1998)}]{Tully1998Mixed}%
  \BibitemOpen
  \bibfield  {author} {\bibinfo {author} {\bibfnamefont {J.}~\bibnamefont {Tully}},\ }\href {\doibase 10.1039/A801824C} {\bibfield  {journal} {\bibinfo  {journal} {Faraday Discussions}\ }\textbf {\bibinfo {volume} {110}},\ \bibinfo {pages} {407} (\bibinfo {year} {1998})}\BibitemShut {NoStop}%
\bibitem [{\citenamefont {Parandekar}\ and\ \citenamefont {Tully}(2005)}]{Parandekar2005Mixed}%
  \BibitemOpen
  \bibfield  {author} {\bibinfo {author} {\bibfnamefont {P.~V.}\ \bibnamefont {Parandekar}}\ and\ \bibinfo {author} {\bibfnamefont {J.}~\bibnamefont {Tully}},\ }\href {\doibase 10.1063/1.1856460} {\bibfield  {journal} {\bibinfo  {journal} {The Journal of chemical physics}\ }\textbf {\bibinfo {volume} {122 9}},\ \bibinfo {pages} {094102} (\bibinfo {year} {2005})}\BibitemShut {NoStop}%
\bibitem [{\citenamefont {Castro}\ and\ \citenamefont {Gross}(2013)}]{Castro2013Optimal}%
  \BibitemOpen
  \bibfield  {author} {\bibinfo {author} {\bibfnamefont {A.}~\bibnamefont {Castro}}\ and\ \bibinfo {author} {\bibfnamefont {E.}~\bibnamefont {Gross}},\ }\href {\doibase 10.1088/1751-8113/47/2/025204} {\bibfield  {journal} {\bibinfo  {journal} {Journal of Physics A: Mathematical and Theoretical}\ }\textbf {\bibinfo {volume} {47}} (\bibinfo {year} {2013}),\ 10.1088/1751-8113/47/2/025204}\BibitemShut {NoStop}%
\bibitem [{\citenamefont {Todorov}(2001)}]{todorov2001time}%
  \BibitemOpen
  \bibfield  {author} {\bibinfo {author} {\bibfnamefont {T.}~\bibnamefont {Todorov}},\ }\href@noop {} {\bibfield  {journal} {\bibinfo  {journal} {Journal of Physics: Condensed Matter}\ }\textbf {\bibinfo {volume} {13}},\ \bibinfo {pages} {10125} (\bibinfo {year} {2001})}\BibitemShut {NoStop}%
\bibitem [{\citenamefont {Halliday}\ and\ \citenamefont {Artacho}(2022)}]{halliday2022manifold}%
  \BibitemOpen
  \bibfield  {author} {\bibinfo {author} {\bibfnamefont {J.}~\bibnamefont {Halliday}}\ and\ \bibinfo {author} {\bibfnamefont {E.}~\bibnamefont {Artacho}},\ }\href@noop {} {\bibfield  {journal} {\bibinfo  {journal} {SciPost Physics}\ }\textbf {\bibinfo {volume} {12}},\ \bibinfo {pages} {020} (\bibinfo {year} {2022})}\BibitemShut {NoStop}%
\bibitem [{\citenamefont {De~Aguiar}(2010)}]{de2010intersections}%
  \BibitemOpen
  \bibfield  {author} {\bibinfo {author} {\bibfnamefont {F.}~\bibnamefont {De~Aguiar}},\ }\href@noop {} {\bibfield  {journal} {\bibinfo  {journal} {Physics Letters A}\ }\textbf {\bibinfo {volume} {374}},\ \bibinfo {pages} {4297} (\bibinfo {year} {2010})}\BibitemShut {NoStop}%
\bibitem [{\citenamefont {Urru}\ \emph {et~al.}(2015)\citenamefont {Urru}, \citenamefont {Cocco},\ and\ \citenamefont {Fiorentini}}]{urru2015tunability}%
  \BibitemOpen
  \bibfield  {author} {\bibinfo {author} {\bibfnamefont {A.}~\bibnamefont {Urru}}, \bibinfo {author} {\bibfnamefont {G.}~\bibnamefont {Cocco}}, \ and\ \bibinfo {author} {\bibfnamefont {V.}~\bibnamefont {Fiorentini}},\ }\href@noop {} {\bibfield  {journal} {\bibinfo  {journal} {arXiv preprint arXiv:1511.01341}\ } (\bibinfo {year} {2015})}\BibitemShut {NoStop}%
\bibitem [{\citenamefont {Oh}\ \emph {et~al.}(2008)\citenamefont {Oh}, \citenamefont {Huang}, \citenamefont {Peskin},\ and\ \citenamefont {Kais}}]{oh2008entanglement}%
  \BibitemOpen
  \bibfield  {author} {\bibinfo {author} {\bibfnamefont {S.}~\bibnamefont {Oh}}, \bibinfo {author} {\bibfnamefont {Z.}~\bibnamefont {Huang}}, \bibinfo {author} {\bibfnamefont {U.}~\bibnamefont {Peskin}}, \ and\ \bibinfo {author} {\bibfnamefont {S.}~\bibnamefont {Kais}},\ }\href@noop {} {\bibfield  {journal} {\bibinfo  {journal} {Physical Review A—Atomic, Molecular, and Optical Physics}\ }\textbf {\bibinfo {volume} {78}},\ \bibinfo {pages} {062106} (\bibinfo {year} {2008})}\BibitemShut {NoStop}%
\bibitem [{\citenamefont {Mead}\ and\ \citenamefont {Truhlar}(1979)}]{mead1979determination}%
  \BibitemOpen
  \bibfield  {author} {\bibinfo {author} {\bibfnamefont {C.~A.}\ \bibnamefont {Mead}}\ and\ \bibinfo {author} {\bibfnamefont {D.~G.}\ \bibnamefont {Truhlar}},\ }\href@noop {} {\bibfield  {journal} {\bibinfo  {journal} {The Journal of Chemical Physics}\ }\textbf {\bibinfo {volume} {70}},\ \bibinfo {pages} {2284} (\bibinfo {year} {1979})}\BibitemShut {NoStop}%
\bibitem [{\citenamefont {Akimov}(2018)}]{akimov2018simple}%
  \BibitemOpen
  \bibfield  {author} {\bibinfo {author} {\bibfnamefont {A.~V.}\ \bibnamefont {Akimov}},\ }\href@noop {} {\bibfield  {journal} {\bibinfo  {journal} {The journal of physical chemistry letters}\ }\textbf {\bibinfo {volume} {9}},\ \bibinfo {pages} {6096} (\bibinfo {year} {2018})}\BibitemShut {NoStop}%
\bibitem [{\citenamefont {Mozhayskiy}\ \emph {et~al.}(2006)\citenamefont {Mozhayskiy}, \citenamefont {Babikov},\ and\ \citenamefont {Krylov}}]{mozhayskiy2006conical}%
  \BibitemOpen
  \bibfield  {author} {\bibinfo {author} {\bibfnamefont {V.~A.}\ \bibnamefont {Mozhayskiy}}, \bibinfo {author} {\bibfnamefont {D.}~\bibnamefont {Babikov}}, \ and\ \bibinfo {author} {\bibfnamefont {A.~I.}\ \bibnamefont {Krylov}},\ }\href@noop {} {\bibfield  {journal} {\bibinfo  {journal} {The Journal of chemical physics}\ }\textbf {\bibinfo {volume} {124}} (\bibinfo {year} {2006})}\BibitemShut {NoStop}%
\end{thebibliography}%

\end{document}